# Piezoresistive PtSe$_2$ pressure sensors with reliable high sensitivity and their integration into CMOS ASIC substrates

Sebastian Lukas[1], Nico Rademacher[1,2], Sofía Cruces[1], Michael Gross[1], Eva Desgué[3], Stefan Heiserer[4], Nikolas Dominik[4], Maximilian Prechtl[4], Oliver Hartwig[4], Cormac Ó Coileáin[4], Tanja Stimpel-Lindner[4], Pierre Legagneux[3], Arto Rantala[5], Juha-Matti Saari[5], Miika Soikkeli[5], Georg S. Duesberg[4], Max C. Lemme[1,2]

[1]Chair of Electronic Devices, RWTH Aachen University, Otto-Blumenthal-Str. 25, 52074 Aachen, Germany

[2]AMO GmbH, Advanced Microelectronic Center Aachen, Otto-Blumenthal-Str. 25, 52074 Aachen, Germany

[3]THALES R&T, 1 Av. Augustin Fresnel, 91767 Palaiseau, France

[4]Institute of Physics & SENS Research Centre, University of the Bundeswehr Munich, Werner-Heisenberg-Weg 39, 85577 Neubiberg, Germany

[5]VTT Technical Research Centre of Finland Ltd, P.O. Box 1000, FI-02044 VTT, Espoo, Finland

## Abstract

Membrane-based sensors are an important market for microelectromechanical systems (MEMS). Two-dimensional (2D) materials, with their low mass, are excellent candidates for suspended membranes to provide high sensitivity, small footprint sensors. The present work demonstrates pressure sensors employing large-scale-synthesized 2D platinum diselenide (PtSe$_2$) films as piezoresistive membranes supported only by a thin polymer layer. We investigate three different synthesis methods with contrasting growth parameters and establish a reliable high yield fabrication process for suspended PtSe$_2$/PMMA membranes across sealed cavities. The pressure sensors reproducibly display sensitivities above $6 \cdot 10^{-4}$ kPa$^{-1}$. We show that the sensitivity clearly depends on the membrane diameter and the piezoresistive gauge factor of the PtSe$_2$ film. Reducing the total device size by decreasing the number of membranes within a device leads to a significant increase in the area-normalized sensitivity. This allows the manufacturing of pressure sensors with high sensitivity but a much smaller device footprint than the current state-of-the-art MEMS technology. We further integrate PtSe$_2$ pressure sensors with CMOS technology, improving the technological readiness of PtSe$_2$-based MEMS and NEMS devices.

## Introduction

Research on micro- and nanoelectromechanical systems (MEMS and NEMS) has increasingly included two-dimensional (2D) materials. Their crystalline, layered nature promises decisive advantages, especially in the area of sensing [1–3]. One outstanding property of several 2D materials is their piezoresistivity, which can be more fully exploited because of their atomic thinness. Piezoresistive sensors utilizing 2D materials have been demonstrated in the form of graphene-based pressure sensors [4–14] and accelerometers [15], as well as pressure [16] and mass sensors [17] made from one of the best-studied 2D materials, molybdenum disulfide ($MoS_2$). Noble-metal dichalcogenides (NMDCs), a subgroup of transition-metal dichalcogenides (TMDCs) in which the metal is a noble metal such as platinum (Pt) or palladium (Pd), have also recently gained attention. The piezoresistive properties of platinum diselenide ($PtSe_2$, shown with its atomic structure in Figure 1a) have been studied [18,19], and $PtSe_2$-based piezoresistive pressure sensors have been demonstrated [20,21]. Although the resonance [22,23] and capacitance [24–28] of 2D membranes have also been used as pressure sensor mechanisms, the piezoresistivity of 2D materials is a simple transduction mechanism for sensors with high sensitivity. A piezoresistive pressure sensor based on the NMDC palladium diselenide ($PdSe_2$) has been shown to reach a sensitivity of $3.9 \cdot 10^{-4}$ $kPa^{-1}$; however, a silicon nitride ($SiN_x$) membrane of very large dimensions was used as a mechanical support [29]. Utilizing the 2D material itself as a membrane instead would truly exploit the advantages of 2D materials in membrane-based sensors.

In our study, we aim to approach the ultimate limit of 2D membrane-based sensors. We used nanocrystalline layered $PtSe_2$ from three different synthesis methods: thermally assisted conversion (TAC) of predeposited Pt layers [30–33], metal–organic chemical vapor deposition (MOCVD) [34,35], and molecular beam epitaxy (MBE) [36–38]. These three synthesis methods each have their own advantages and disadvantages. While MBE and MOCVD are more likely to result in well-aligned $PtSe_2$ layers with larger crystallites (grains), TAC is very easily scalable to industry-relevant wafer sizes and can simultaneously produce high-quality $PtSe_2$ films at low temperatures (< 450 °C), compatible with complementary metal-oxide-semiconductor (CMOS) back-end-of-line thermal budgets. Based on our knowledge of the process of fabricating suspended 2D material membranes on sealed cavity substrates [21,39,40], we constructed and characterized pressure sensors from a total of 13 different growth batches with different synthesis parameters. The suspended $PtSe_2$ membranes were examined via atomic force microscopy (AFM) and Raman spectroscopy depth scans. The pressure sensors were measured in an automated pressure chamber, and their sensitivity was extracted. In parallel, the piezoresistive gauge factor (GF) was determined from $PtSe_2$ samples of the same growth batches, and its influence on the sensor sensitivity is shown. Finally, we also present $PtSe_2$ pressure sensors fabricated via the same methods on top of CMOS application-specific integrated circuit (ASIC) substrates from a semiconductor foundry. This demonstrates that our technology can be implemented in well-established CMOS fabrication lines.

**Results**

PtSe$_2$/poly(methyl methacrylate) (PMMA) membrane-based pressure sensors with closed cavities and bottom contacts were fabricated as described in the Methods section. PtSe$_2$ synthesized by TAC, MOCVD, or MBE was transferred to target substrates and thereby suspended above pre-etched cavities via a frame-based transfer process and a subsequent patterning step. A schematic of the device cross section is shown in Figure 1b together with the process flow. A summary of the fabricated and analyzed chips is shown in Table 1.

Each chip had up to 48 sensor devices depending on the area covered with PtSe$_2$ during transfer, with six different membrane diameters ranging from 1.5 µm to 12 µm. Each of the devices comprised arrays of these suspended PtSe$_2$/PMMA membranes. The yield of intact membranes after transfer and patterning (lithography, etching, and resist removal) was generally high. Broken membranes occurred only in places where small bubbles or wrinkles had formed during transfer or where particles prevented good adhesion of PtSe$_2$ to SiO$_2$. In those places, liquid chemicals (developer, solvent) were able to penetrate the interface and delaminate the PtSe$_2$. Nevertheless, many devices with very high yields of intact membranes were found across all the chips.

Optical microscope images of the fabricated devices are shown in Figure 1d-e and Figure S1. The color difference between chips originates from factors such as the difference in PtSe$_2$ thickness in the channel areas and overetching during the patterning process in the surrounding areas of the SiO$_2$ substrate.

Atomic force microscopy (AFM) confirmed the suspended nature of the PtSe$_2$/PMMA membranes above the void, with 2.0-2.5 µm deep cavities, as shown in Figure 2a and Figure S2. The membranes have a rather flat topography, with a maximum deflection of approximately 50 nm across a scanned membrane diameter of 5.2 µm, as depicted in Figure 2b. The difference in the deflection and surface morphology between the chips can be attributed to small changes in the manual transfer process, differences in the adhesion of PtSe$_2$ to its growth substrate and therefore different applied tensions during delamination, and differences in the nanocrystalline structure of the various PtSe$_2$ films.

Additionally, Raman depth scanning [41] was performed on devices with 12 µm membrane diameters. This method uses several line scans with consecutive laser focus planes in the z-direction to reconstruct two-dimensional images of the device cross sections (see Figure 2c and Figure S3). The intensity maps of the PtSe$_2$ Raman peak (E$_g$ peak at ~178 rel. cm$^{-1}$) show that the PtSe$_2$/PMMA membranes are freely suspended, whereas the intensity maps of the silicon (Si) Raman peak display the profile of the 2.0-2.5 µm deep cavities below.

Four-probe back-gated transfer measurements were conducted using six-port reference devices without membranes fabricated on the same substrates as the pressure sensors. The field-effect charge carrier mobility $\mu_\text{F}$ and sheet resistance $R_\text{sheet}$ of the transferred PtSe$_2$ films were determined as described in the Methods section and are displayed in Figure S4. Most PtSe$_2$ films have $R_\text{sheet}$ values in the range of a few kΩ to a few 100 kΩ and a maximum $\mu_\text{F}$ between 1 and 12 cm²/Vs. One chip of MOCVD-PtSe$_2$ (#266.4) has a higher $R_\text{sheet}$ of approximately 10 MΩ up to 1 GΩ and, at the same time, a very low $\mu_\text{F}$.

The piezoresistive gauge factor (GF) was determined for all different batches of PtSe$_2$ films (except for one MBE film, for which no additional material from the same batch was available). The PtSe$_2$ was exposed to a defined strain, and the electrical resistance was measured, as described in the Methods section and shown in Figure S5, Figure S6, and Figure S7. The extracted GFs are listed in Table 1. The GF varies between the samples, ranging from -55.6 to +4.2, while most values are close to approximately -35. The origin of the spread of the GF is related to the layer thickness as well as the nanocrystalline structure of the PtSe$_2$ and is difficult to predict [18,19]. The GF is an important indicator of the resulting sensitivity of the membrane-based pressure sensor.

The wire-bonded pressure sensor chips were characterized in an automated pressure chamber with various gas pressures between 200 mbar and 1550 mbar (see Figure 3a-b). The pressure, temperature, and relative humidity were monitored by commercial reference sensors inside the chamber. For all the measurements, the chamber was purged with nitrogen to ensure a low relative humidity (< 15 %) and therefore minimize the reaction of the sensor to a change in relative humidity. The temperature was stable between 22 and 25 °C. The chips were exposed to ambient light during the measurements, which, as verified in a reference measurement (see Figure S8a), did not lead to significant additional drifts of the sensor output owing to its low intensity.

The device resistance was recorded while the pressure inside the chamber was cycled between low and high pressures (see Figure 3d-f as well as Figure S9 and Figure S10). Owing to the difference in the pressure inside the cavities sealed by the PtSe$_2$ membranes, $p_\text{internal}$, and the external pressure $p_\text{external}$, the membrane is deflected and thereby strained. A schematic of this operation principle is shown in Figure 3g. The strain leads to a change in the electric resistance of PtSe$_2$, $\Delta R \propto GF \cdot \varepsilon$. The signal shape of the devices varies depending on several factors, mainly the membrane diameter of the device and the PMMA thickness, as previously analyzed and simulated [21]. In this study, the PMMA thickness was kept the same for all the fabricated and measured chips. We used approximately 60 nm of PMMA A2 950k, which was selected as a compromise between sufficient mechanical stability for transfer and sufficient thinness to ensure the high sensitivity of the sensors.

For a PMMA thickness of approximately 60 nm, the sensors exhibit two clearly distinguishable signal shapes, both of which are shown in Figure 3e-f. The device in Figure 3e has comparably large membranes with diameters of 12 µm. The PtSe$_2$ in this device dominantly experiences tensile strain in both directions of deflection (outward or inward for low and high pressures, respectively). Therefore, resistance minima can be observed for both high- and low-pressure cases in the $R$ curve of this device, where the tensile strain is maximized and, owing to the negative GF, the device resistance is minimized. The maxima in $R$ are observed at a pressure of approximately 835 mbar, corresponding to the minima in tensile strain, i.e., a non-deflected, flat membrane. The exact pressure of the resistance maximum varies from device to device and depends on the pressure history. At an ambient pressure of approximately 1050 mbar, the membranes are therefore not unstrained but have already deflected inward. The different cases are shown in the schematics in Figure 3h-k. For the devices exhibiting the described signal shape, the sensitivity was extracted between the low-pressure level and the flat-membrane resistance maximum, as well as between the mentioned resistance maximum and the high-pressure level (as indicated in Figure 3e).

A different signal shape was observed for several devices with a smaller membrane diameter, such as the device shown in Figure 3f, with a membrane diameter of 3.4 µm. Here, the piezoresistively dominating type of strain changes from dominating tensile strain for the low-pressure case to a combination of tensile and compressive strain across each membrane (with a possible slight domination of compressive strain) for the high-pressure case. Compressive strain occurs in the PtSe$_2$ film due to its position at the bottom of the double-layer stack of PtSe$_2$ and PMMA, where the PMMA is approximately 3 to 18 times as thick as the PtSe$_2$. The dominating strain occurs around the edge of the membrane, as shown previously [21,42]. As a result, the measured devices show a resistance minimum for the low-pressure case and a resistance maximum for the high-pressure case, between which the sensitivity was extracted.

The extracted sensitivity is plotted against the membrane diameter in Figure 4a-b. The error bars show the ranges of the extracted sensitivities for each chip. A total of 264 devices across the 17 different chips with PtSe$_2$ from 13 different growth batches (i.e., various growth methods and/or growth parameters) were characterized via the main device layout. Additionally, 14 devices without membranes were measured as reference devices and showed only a minor response with very low sensitivity, approximately two orders of magnitude lower than the sensitivity extracted from those devices with membranes, which was attributed to small changes in relative humidity (see four exemplary curves in Figure S11). These reference devices are plotted with a membrane diameter of zero in Figure 4a-b.

Generally, a trend of increasing sensitivity with increasing membrane diameter can be observed. This is expected due to the increase in strain with increasing membrane diameter [21]. Parabolic fits are calculated and plotted for all different chips, following the quadratic dependence of the membrane area on its diameter. However, multiple factors influence the sensitivity of the sensors in addition to the membrane diameter, such as the total active area (i.e., the area of the channel that is covered by intact membranes; see also Table S1) and processing variations, which cause differences in the pretension of the membranes. The parabolic fit nevertheless matches the data well for many chips.

One chip (#151.2) shows an opposite signal shape; therefore, a sensitivity with the opposite sign is extracted (see Figure S9c and Figure 4a). This can be explained by the GF of this $PtSe_2$ film, which, in contrast to all other samples, has a positive sign. The dependence of the sensitivity on the GF is further shown in Figure 4c, where the sensitivity values of all the measured devices with the three largest membrane diameters are plotted against the GF. Despite the significant spread of the data, linear fits visualize the clear trends of higher sensitivities with larger membrane diameters as well as higher sensitivities with larger GF magnitudes.

The device design with a large $PtSe_2$ channel and hundreds to thousands of membranes in each single device (see Table S1) has the disadvantage of low sensitivity normalized to the membrane area $S_A = S/A_{\text{membranes,total}}$ but was chosen earlier when the transfer process still resulted in lower yields of intact membranes. To show that a much higher $S_A$ is possible with the demonstrated technology, one chip (#216.3) was fabricated with enlarged Ti/Pd contact pads, covering significant parts of the cavity array and therefore reducing the active membrane area of the devices. Details on this chip can be found in Figure S12. The extracted sensitivity $S$ of the two measured devices with reduced active membrane areas was on the same order of magnitude as that of the previously characterized chips (see the turquoise-colored stars in Figure 4a), and the extracted sensitivity normalized to the membrane area ($S_A$) was greater by a factor of 5 to 6 than that of the chips with $PtSe_2$ from the same batch. For future device designs (and as already considered in the later-designed devices on CMOS ASIC substrates, see below), a much smaller footprint with fewer membranes per device should therefore be considered to achieve an even greater $S_A$.

The nonlinearity (NL) of the PtSe$_2$ pressure sensors was evaluated for all the measured devices. The maximum $R$, corresponding to the flat membrane state of minimum strain, leads to a very nonlinear response across the whole pressure range from low to high pressure. Therefore, the NL was calculated for two ranges for each device, from low pressure to 50 mbar below the flat membrane pressure and from 50 mbar above the flat membrane pressure to high pressure, according to $NL(p) = (R_{p,\text{measured}} - R_{p,\text{linear}})/(R_{p,\text{max}} - R_{p,\text{min}})$, where $R_{p,\text{measured}}$ is the measured sensor resistance at pressure $p = (p_{\text{max}} - p_{\text{min}})/2$, between the resistances $R_{p,\text{max}}$ and $R_{p,\text{min}}$ at the upper and lower bounds of the respective range, $p_{\text{max}}$ and $p_{\text{min}}$. $R_{p,\text{linear}}$ is the theoretical resistance of the sensor at pressure $p$, assuming a linear relationship of $R$ and $p$ between $R_{p,\text{max}}$ and $R_{p,\text{min}}$ [43,44]. $R$-$p$ plots of several devices are shown in Figure S13a-d, with the ranges for NL extraction indicated. The NL is dependent on the membrane diameter $d$ (see Figure 4d), with an increase in $d$ leading to a better (i.e., lower) NL. Almost all PtSe$_2$ pressure sensors with $d = 12$ µm reach an NL below 4 %, whereas several even have an NL below 1 % in both pressure ranges (see Figure 4e and, for example, chip #216.3 with a reduced active membrane area in Figure S13a).

Long-term measurements were performed on two devices on chip #220.1 (see Figure S8b), which previously demonstrated high sensitivity and rather little drift. Starting from ambient pressure, the pressure was decreased to 200 mbar and held constant for more than 16 minutes. After several minutes at 200 mbar, a drift of the device resistance toward higher values was observed, which can be attributed to slow leakage of the gas encapsulated below the membranes and, therefore, a slow reduction in the membrane deflection and strain. In subsequent measurements, the pressure at which the device resistance was maximized shifted to lower pressures, indicating that the flat membrane state was now reached at a lower pressure, corresponding to the pressure inside the cavities. Nevertheless, the drift is not significant on the order of seconds. In fact, a very slow drift of a pressure sensor back to its most sensitive configuration under permanent changes in ambient conditions might be favorable for certain applications.

In addition to the devices discussed to this point, pressure sensors were also fabricated on top of commercially fabricated dedicated CMOS ASIC substrates with read-out circuitry (including multiplexers for device selection) in the layers below the pressure sensors. Microscope images of two dies on a CMOS ASIC substrate, a matrix with 16 devices, and a single device are shown in Figure 5a, b, and c, respectively. The measured devices had membrane diameters of 11.6 µm and 13.5 µm. The measurement curves of the two devices are shown in Figure 5d-f, while more data can be found in Figure S14. The average extracted $S$ ranged between $3.1 \cdot 10^{-4}$ kPa$^{-1}$ and $4.7 \cdot 10^{-4}$ kPa$^{-}$1, i.e., in the same range as the devices on non-CMOS substrates. The devices on the CMOS ASIC substrates were designed with smaller membrane arrays due to space restrictions, leading to a much higher $S_A$ (see Figure 4f and Table 2).

**Conclusions**

The present work shows that large-scale synthesized PtSe$_2$ possesses great potential as a piezoresistive membrane material in pressure sensors. A large dataset of devices with PtSe$_2$ from three different synthesis methods, each featuring several synthesis batches with differing parameters, was analyzed. A fabrication process to reliably fabricate suspended PtSe$_2$/PMMA membranes across sealed cavities with a high yield was established. The measured pressure sensors reproducibly showed sensitivities of up to higher than $6 \cdot 10^{-4}$ kPa$^{-1}$ and low nonlinearities of down to < 1 % in several devices. The sensitivity clearly depends on the membrane diameter and the piezoresistive gauge factor of the PtSe$_2$ film. We have demonstrated that a reduction in the device size through a decrease in the number of membranes within a device is possible, leading to a significant increase in the area-normalized sensitivity. This will allow the manufacturing of pressure sensors with high sensitivity but a much smaller device footprint than the current state-of-the-art MEMS technology. We have further demonstrated the feasibility of integrating PtSe$_2$ pressure sensors with CMOS technology, improving the technology readiness of PtSe$_2$-based MEMS and NEMS technology.

# Methods

## Substrate fabrication

Before PtSe$_2$ transfer, target substrates based on 150 mm p-doped Si wafers were prepared. For the marker layer, stepper lithography was performed, and the markers were etched by reactive ion etching (RIE) with C$_4$F$_8$ and SF$_6$. After a second step of stepper lithography for the cavity layer, 2.0–2.5 µm deep cavities were etched into the Si via RIE, again with C$_4$F$_8$ and SF$_6$ chemistry. After a subsequent cleaning step in a mixture of H$_2$O$_2$ and H$_2$SO$_4$, the wafers were thermally oxidized to grow 90 nm of SiO$_2$ on top of the Si for passivation. Oxide growth after etching allows passivating cavity sidewalls and floors, erasing the possibility of short circuits from collapsed PtSe$_2$ membranes later.

Stepper lithography with a resist double layer of lift-off resist (LOR) 3A and the photoactive AZ5214E resist was executed to deposit and pattern bottom contacts consisting of evaporated titanium (Ti, 5 nm) and palladium (Pd, 10 nm). Even though the wafer surface was uneven due to the etched cavities, the metal lift-off was successful because of the resist stack. A final stepper lithography and lift-off step was performed to evaporate and pattern 300 nm thick aluminum (Al) pads on top of the previously patterned metal contacts in the probing pad areas for later wire bonding. Finally, the wafers were diced into 2 cm square chips.

Owing to adhesion issues with the Al pads on the Pd surface in the final step of resist stripping with the TMAH-containing developer after transfer and patterning (see below), several chips were alternatively fabricated with gold (Au) bond pads. In this case, the abovementioned Ti/Pd deposition and subsequent gold (Au) bond pad deposition, instead of Al deposition, were performed via contact lithography on several 2 cm pieces diced from the wafers after thermal oxide growth. The metal stack used for each respective chip is indicated in Table 1. Note that in all the chips, the metal in contact with the PtSe$_2$ was Pd.

For one chip (#216.3), lithography for Ti/Pd metallization was performed with the help of a maskless aligner (Microtech LW405C), extending the metal contacts to cover significant parts of the cavity array and therefore decreasing the active membrane area. The Au bond pads were subsequently deposited via contact lithography.

## CMOS ASIC substrate fabrication

The CMOS wafers used for integration of the PtSe$_2$ pressure sensors were manufactured via standard commercial technology provided by XFAB. The 200 mm CMOS wafers utilize a 0.18 µm analog CMOS process node with up to six metallization layers. The CMOS technology includes a wide range of active devices and high-performance analog devices.

Wafer-scale post-processing was started by etching the vias to the topmost CMOS metal layer by inductively coupled plasma etching (ICP-RIE). Titanium tungsten (TiW)/Au (10 nm/100 nm) via metallization was deposited by sputtering and patterned in a wet etching process. In the next step, Ti/Au (2 nm/30 nm) was deposited by lift-off for the bottom contact metals. The wafers were then diced into single CMOS microsystem chips. All post-process patterning steps were performed via standard UV lithography.

### $PtSe_2$ growth

We used and compared $PtSe_2$ from various synthesis methods.

*Thermally assisted conversion (TAC).* Thin Pt films were initially sputtered onto $Si/SiO_2$ substrates using a Cressington magnetron sputter coater. These films were then converted by TAC to $PtSe_2$ in a custom-built cold wall TAC and CVD system. As a precursor, selenium (Se) powder (> 99.5 % purity, VWR) was heated to 228 °C and transported to the samples by a carrier gas flow, consisting of varying amounts of $N_2$ and $H_2$ in the range of 50-500 sccm. The conversion was carried out for 120 min at a substrate temperature of 450 °C and at a low pressure of < 10 mbar. Samples #184 and #190 were converted at a higher temperature of 550 °C. After conversion, the $PtSe_2$ films tended to be approximately 2.5 to 3 times thicker than the initial Pt layer.

*Molecular beam epitaxy (MBE).* MBE-$PtSe_2$ samples were synthesized in a 2-inch MBE reactor supplied by Dr. Eberl MBE-Komponenten GmbH (Germany). Sapphire (0001) (c-plane) substrates were purchased from Kyocera Corporation (Japan). The sapphire was first chemically cleaned with acetone/propanol, and then a 100-nm-thick molybdenum film was deposited on the back to improve radiative heating. It was outgassed under ultrahigh vacuum ($10^{-10}$ mbar) at 500 °C overnight and then annealed at 900 °C for 15 min. A Se flux was supplied by a valved Se cracker source filled with ultrahigh purity (7N) Se, which was heated to 290 °C, and the cracker temperature was fixed at 600 °C. A Pt flux originated from high-purity (4N) Pt heated in an electron beam evaporator. During synthesis, the sapphire (0001) substrate was heated to the desired growth temperature (463 °C and 520 °C for samples #192 and #195, respectively) and simultaneously exposed to Pt flux (0.003 Å/s) and Se flux (0.2 Å/s and 0.5 Å/s for samples #192 and #195, respectively). Sample #195 was annealed after growth at 690 °C for 30 min under the same Se flux as that used during growth. After synthesis, the substrate was cooled to 200 °C within 20 min continuously under Se flux.

*Metal-organic chemical vapor deposition (MOCVD).* MOCVD growth of $PtSe_2$ was carried out within a custom-built cold-wall reactor [34,35]. Cleaned 20 mm × 20 mm Si chips with a 90 nm $SiO_2$ layer were placed on the substrate plate and heated to 600 °C under low-pressure conditions. Se powder (> 99.5 % purity, VWR) was evaporated at 225 °C and carried toward the reaction chamber by a $H_2$ flow. (Trimethyl)methylcyclopentadienylplatinum(IV) (($CH_3)_3(CH_3C_5H_4$)Pt) (Strem Chemicals, 99 % purity) was heated to 40 °C and used as a metal-organic Pt precursor. The layer thickness was controlled by the process time, which was approximately 10-to 15 min for $PtSe_2$ thicknesses of approximately 3-5 nm.

## $PtSe_2$ transfer

The various $PtSe_2$ films were transferred to the target substrates via a frame-based temperature-assisted dry transfer method, as previously demonstrated for graphene [40]. A very thin layer of approximately 60 nm PMMA A2 950k was spin-coated onto the $PtSe_2$ films on their growth substrates. After baking at 180 °C, frames of transparent plastic foil and heat-resistant polyimide tape were prepared and attached to the edges of the PMMA surface on the growth substrates. Here, polyimide tape with acrylic adhesive (instead of standard silicone adhesive) was used to enhance the chemical stability of the transfer frame in a potassium hydroxide (KOH) solution. In the next step, each transfer frame with the attached PMMA/$PtSe_2$/growth substrate was placed inside a beaker filled with cured polydimethylsiloxane (PDMS), to which the back side of the growth substrate firmly adhered. A pipette was used to push a few milliliters of a 4 mol/L KOH solution below the transfer frame to the edge of the growth substrate. Intercalation of the KOH solution between the $PtSe_2$ and the growth substrate eventually entirely detached the transfer frame with $PtSe_2$/PMMA from the growth substrate (see Figure 1c). After careful removal from the KOH solution, the frame was rinsed in DI water thoroughly and finally placed in air to dry.

The $PtSe_2$/PMMA was then placed on top of a target substrate with its transfer frame on a hotplate at 95 °C, and the adhesive was attached to prevent any sidewards motion of the transferred frame. The temperature of the hotplate was then increased to approximately 160 °C to soften the PMMA above its glass transition temperature and therefore allow the $PtSe_2$ to adhere smoothly to the target substrate. After successful complete adhesion, the $PtSe_2$/PMMA was carefully cut along the inner edge of the transfer frame with a sharp scalpel, and the frame was subsequently removed. The hotplate temperature was increased to 180 °C, and the samples were heated for 10 minutes.

## $PtSe_2$ patterning

$PtSe_2$/PMMA was patterned into the device channel shapes on all samples at the 2 cm chip level via contact lithography. The PMMA was not removed to mechanically support the $PtSe_2$ membranes during the fabrication process or later device operation. LOR 3A was used between PMMA and the photoactive AZ5214E resist to prevent the mixing of the AZ5214E resist and PMMA and therefore the formation of an insoluble layer at their interface. The bottom-to-top layer stack during the lithography step was, therefore, PtSe$_2$, PMMA, LOR 3A, and AZ5214E. After exposure, the AZ was used for development. It does not contain TMAH and, therefore, does not lead to delamination or corrosion of the Al pads. It also does not dissolve LOR 3A. LOR 3A, PMMA and PtSe$_2$ were then etched by a low-power CF$_4$/O$_2$ plasma during RIE. After etching, a low-power O$_2$ plasma treatment was used to remove any cross-linked polymer from the top of the photoresist layer that could otherwise remain after removal in solvents. Finally, on the chips with Al bond pads, the AZ5214E photoresist and LOR were removed by using DMSO at room temperature. Alternatively, in the case of substrates with Au bond pads and in the case of CMOS ASIC substrates, flood exposure and a 30 s dipping in a TMAH-containing developer were used to remove the LOR and AZ5241E photoresist after etching and O$_2$ plasma treatment. This avoided the use of DMSO, which in some cases led to thinning or partial removal of the PMMA layer and therefore to the collapse of the membranes. Optical microscope images of the final devices before wire bonding are shown in Figure 1d-e and Figure S1, as well as in Figure 5 (devices on CMOS ASIC substrates). The channel area on each fabricated device of the in-house-fabricated substrates, comprising the total membrane array area plus the surrounding passive PtSe$_2$ area, was 320 µm × 320 µm and included between 585 and 12,257 individual membranes, depending on the membrane diameter (see Table S1).

*Piezoresistive characterization*

To determine the piezoresistive gauge factor (GF) of the various PtSe$_2$ films, small pieces of the same PtSe$_2$ films used for sensor fabrication were transferred onto prepared flexible polyimide (Kapton) substrates, bridging a few-mm gap between the nickel (Ni) contact pads. The samples were glued to a steel beam, and wires were soldered to the Ni contacts. The steel beam was then bent in a controlled way by applying a weight so that a tensile or compressive strain of $\varepsilon = 0.0044$ % was applied to the sample. The setup is described in more detail in refs. [18,20]. A Keithley 4200-SCS parameter analyzer was used for sampling the strain gauge resistance $R$ while repeatedly applying and removing the attached mass over several minutes. The piezoresistive gauge factor GF was then extracted via $\mathrm{GF} = \Delta R/(R_0 \cdot \varepsilon)$ [45], where $\Delta R$ is the change in device resistance from the initial value $R_0$.

*Raman spectroscopy*

A WITec alpha300 R Raman microscope with a 532 nm laser and a 1800 g/mm grating was employed for Raman cross-sectional depth scanning [41] of the membrane devices. The focal point of the laser beam had an approximate diameter of 300 nm and a depth of ≤ 1 µm in the direction of the beam. We therefore chose ⅓ µm as the lateral pixel size and set the pixel spacing in the z-direction to 1 µm. A power of only 0.1 mW was used since a power of just 0.3 mW was sufficient to burn through and destroy the suspended PtSe₂/PMMA.

*Atomic force microscopy (AFM)*

AFM scanning of the membrane devices was performed via a Bruker Dimension Icon atomic force microscope in tapping mode. In particular, 15 µm × 15 µm scans of the suspended PMMA/PtSe₂ membranes were recorded at a scan rate of 0.25 Hz.

*Electrical characterization*

A Cascade Summit 12000 semiautomatic probe station connected to a Hewlett Packard 4156B Precision Semiconductor Parameter Analyzer and a Hewlett Packard E5250A Low Leakage Switch Mainframe was used for electrical measurements of the back-gated test structures. Measurements were performed via Keysight WaferPro Express test routine software. The field-effect mobility was calculated according to $\mu_F = (\partial(I_D/V_{\text{diff}})/\partial V_{\text{BG}}) \cdot L_{\text{inner}} \cdot d_{\text{ox}}/(W \cdot \varepsilon_{\text{r,ox}} \cdot \varepsilon_0)$ from the four-probe field-effect measurements on the six-port devices, where $I_D$ is the current along the PtSe₂ channel, $V_{\text{diff}}$ is the voltage between the two inner contacts, $V_{\text{BG}}$ is the back-gate voltage, $L_{\text{inner}} \approx 25$ µm is the distance between the two inner contacts, $d_{\text{ox}} = 90$ nm is the thickness of the gate oxide, $W \approx 10$ µm is the PtSe₂ channel width, and $\varepsilon_{\text{r,ox}} = 3.9$ and $\varepsilon_0$ are the relative permittivity of the gate oxide and the vacuum permittivity, respectively. The precise values for $W$ and $L_{\text{inner}}$ were estimated from the dimensions of the photomask. We also extracted the sheet resistance from the same measurements according to $R_{\text{sheet}} = V_{\text{diff}} \cdot W/(I_D \cdot L_{\text{inner}})$.

The 20 mm chips were manually diced into approximately 6 mm square pieces and then wire bonded into 44-pin ceramic chip carriers (LCCs44) using a 25 µm diameter Au wire with a tpt HB16 wire bonder by ball bonding (see Figure 3a). The chip carriers were then inserted into a pressure chamber setup built in house previously used in refs. [20,40], enabling automation of the valves for pressure control and facilitating synchronized data acquisition of the included reference sensors and the device under test (DUT) (see Figure 3b). The chamber pressure was repeatedly modulated between approximately 200 mbar and 1,500 mbar using a vacuum pump and nitrogen (N₂) gas. A Keithley 4200-SCS parameter analyzer was used for the electrical measurement of the PtSe₂-membrane-based pressure sensors by sampling for several minutes. The sensitivity $S$ was extracted as in $S = \Delta R/(R_0 \cdot \Delta p)$, with the device resistance $R_0$ at ambient pressure $p_0$ and the difference in resistance $\Delta R = R - R_0$ at the applied pressure difference $\Delta p = p - p_0$.

A technology demonstrator (see Figure 3c) was built to show the functionality of the PtSe$_2$ pressure sensor without the need for a pressure chamber or parameter analyzer. The demonstrator houses the pressure sensor chip inside a plastic enclosure glued to a printed circuit board (PCB). The PCB features an Arduino and a display, as well as small switches to select the device to be read. A rubber tube is connected to the plastic enclosure and to a manual 250-mL syringe, allowing the gas pressure to be reduced by pulling the syringe. The sensor resistance is digitally filtered and then displayed. A change in the resistance under a change in pressure can be observed, qualitatively demonstrating the functionality of the sensor. The demonstrator is powered through a USB connection (5 V). Using the Arduino Integrated Development Environment (IDE) on a computer, multiple sensor responses over time can also be displayed in a graph.

The CMOS ASIC substrates were connected to an additional dedicated PCB to address the devices through the CMOS circuitry. The columns and rows of the device to be read were set through a Python script controlling the PCB. The Keithley 4200-SCS parameter analyzer was then connected to the PCB for the measurement.

**Acknowledgments**

This work received funding from the German Ministry of Education and Research (BMBF) under grant agreement 16ES1121 (ForMikro-NobleNEMS), from the European Union's Horizon 2020 research and innovation programme under grant agreement 881603 (Graphene Flagship Core 3), from the European Union's Horizon Europe research and innovation programme under grant agreement 101135196 (2D-PRINTABLE), from the German Research Foundation (DFG) under grant agreement LE 2441/11-1 (2D-NEMS), and from the French ANR agency through the project ANR-20-CE09-0026 "2DonDemand". The authors thank dtec.bw—Digitalization and Technology Research Center of the Bundeswehr for support (project VITAL-SENSE). dtec.bw is funded via the German Recovery and Resilience Plan by the European Union (NextGenerationEU).

**Data Availability**

All the data are available upon reasonable request to the authors.

Table 1. Overview of the fabricated and analyzed pressure sensor chips. Chip #216.3 (marked with *) featured a modified device design with a reduced membrane area. Chips #221.1 and #221.2 (marked with **) were CMOS ASIC substrates featuring a different device design from all other chips.

| source | synthesis method | chip name | approx. $PtSe_2$ thickness | GF | contact metal stack |
|---|---|---|---|---|---|
| University of the Bundeswehr Munich | thermally assisted conversion (TAC) | #133.2 | 12.2 nm | -6.2 | Ti/Pd + Au |
| | | #151.2 | 23.5 nm | +4.2 | Ti/Pd + Au |
| | | #180.3 | 6.1 nm | -33.2 | Ti/Pd + Al |
| | | #182.2 | 4.2 nm | -46.8 | Ti/Pd + Au |
| | | #183.4 | 3.3 nm | -55.6 | Ti/Pd + Au |
| | | #184.3 | 4.2 nm | -47.9 | Ti/Pd + Au |
| | | #190.2 | 14.9 nm | -32.4 | Ti/Pd + Au |
| | | #216.3 * | 3.5 nm | -30.6 | Ti/Pd + Au |
| | | #220.1 | | | Ti/Pd + Al |
| | | #220.3 | | | Ti/Pd + Au |
| | | #220.4 | | | Ti/Pd + Au |
| | | #221.1 ** | | | Ti/Au |
| | | #221.2 ** | | | Ti/Au |
| | | #249.1 | 20 nm | -29.6 | Ti/Pd + Au |
| Thales R&T | molecular beam epitaxy (MBE) | #192.3 | 8.5 nm | unknown | Ti/Pd + Al |
| | | #192.4 | | | Ti/Pd + Au |
| | | #195.8 | 7 nm | -38.2 | Ti/Pd + Au |
| | | #195.9 | | | Ti/Pd + Au |
| University of the Bundeswehr Munich | metal-organic chemical vapor deposition (MOCVD) | #266.4 | 5 nm | -36.0 | Ti/Pd + Au |
| | | #268.6 | 3.5-5 nm | -31.3 | Ti/Pd + Au |

*Table 2. Comparison of the mean sensitivity and the mean area-normalized sensitivity of Si-based and 2D material-based pressure sensors.*

| device structure | mean sensitivity [$10^{-4}$ kPa$^{-1}$] | suspended area [µm$^2$] | mean sensitivity normalized to membrane area [$10^{-8}$ kPa$^{-1}$ µm$^{-2}$] | reference |
|---|---|---|---|---|
| Si membrane with boron-implanted piezoresistors | 0.24 | π·($80^2$-$20^2$) = 18,850 | 0.13 | Godovitsyn et al., 2013 [46] |
| Si membrane with boron-implanted piezoresistors | 0.32 | 450 × 1,000 = 450,000 | 0.0071 | Infineon MultiMEMS platform pressure sensor for tire-pressure monitoring system (TPMS) [43] |
| suspended graphene without PMMA, graphene grown by CVD | 0.3 | 6 × 64 = 384 | 7.7 | Smith et al., 2013 [7] |
| graphene on perforated SiN$_x$ membrane, graphene grown by CVD | 2.8 | 490 × 490 = 240,100 | 0.12 | Wang et al., 2016 [5] |
| BN/graphene/BN, graphene grown by CVD | 1.9 | 6 × 64 = 384 | 49.5 | Li et al., 2018 [11] |
| suspended graphene with 50 nm PMMA, graphene grown by CVD | 0.29 | 20 × π·$5^2$ = 1,571 | 1.83 | Lin et al., 2019 [12] |
| suspended graphene with 50 nm PMMA, graphene grown by CVD | 0.74 | 3 × π·$8^2$ = 603 | 12.3 | Liu et al., 2020 [8] |
| hBN/graphene/hBN, graphene grown by CVD | 2.9 | 210 × 210 = 44,100 | 0.66 | Wang et al., 2022 [13] |
| suspended graphene with 100 nm Si$_3$N$_4$, graphene grown by CVD | 5.3 | 9 × 64 = 576 | 92.4 | Zeng et al., 2023 [14] |
| graphene/hBN/graphene membrane (simulated only) | 4.5 | 6 × 6 = 36 | 1,249 | Zeng et al., 2024 [47] |
| PdSe$_2$ on SiN$_x$ membrane | 3.9 | 700 × 700 = 490,000 | 0.08 | Gong et al., 2024 [29] |
| suspended PtSe$_2$/PMMA, PtSe$_2$ grown by TAC | 13.9 | 157 | 885 | Wagner et al., 2018 [20,42] |
| suspended PtSe$_2$ without PMMA, PtSe$_2$ grown by MBE | 3.8 | 66,162 | 0.57 | Lukas et al., 2023 [21] |
| suspended PtSe$_2$ with 60 nm PMMA, PtSe$_2$ grown by TAC | 6.1 | 66,162 | 0.93 | this work (#180.3, $d_{mem}$ = 12 µm) |
| suspended PtSe$_2$ with 60 nm PMMA, PtSe$_2$ grown by TAC | 3.7 | 8,822 (approx.) | 4.1 | this work (#216.3, $d_{mem}$ = 12 µm, device E1-06) |
| suspended PtSe$_2$ with 60 nm PMMA, PtSe$_2$ grown by TAC | 2.8 | 5,881 (approx.) | 4.8 | this work (#216.3, $d_{mem}$ = 12 µm, device E1-12) |
| suspended PtSe$_2$ with 60 nm PMMA, PtSe$_2$ grown by TAC | 4.7 | 1,268 | 36.9 | this work (#221.1, $d_{mem}$ = 11.6 µm, device C4R2, CMOS ASIC substrate) |
| suspended PtSe$_2$ with 60 nm PMMA, PtSe$_2$ grown by MBE | 4.0 | 66,162 | 0.61 | this work (#192.4, $d_{mem}$ = 12 µm) |
| suspended PtSe$_2$ with 60 nm PMMA, PtSe$_2$ grown by MOCVD | 5.2 | 60,536 | 0.86 | this work (#266.4, $d_{mem}$ = 7.9 µm) |

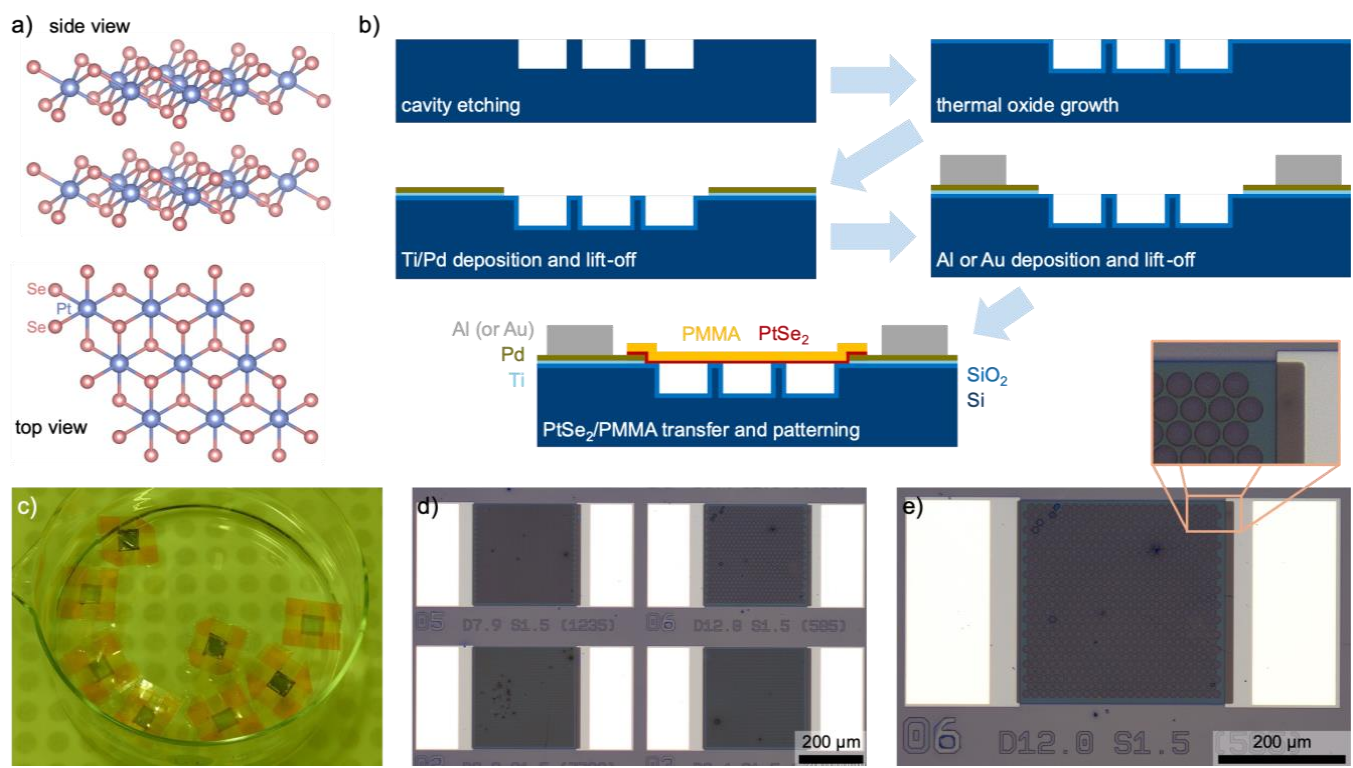

Figure 1. (a) Atomic structure of PtSe$_2$ (side view of two monolayers and top view of a single monolayer). Crystal data from the Materials Project [48,49], graphical representation created using VESTA [50]. (b) Cross-sectional schematics of the process flow for the fabrication of PtSe$_2$/PMMA pressure sensors. (c) Photograph of several PtSe$_2$/PMMA films inside their transfer frames after delamination from the growth substrate. (d-e) Optical microscopy images of PtSe$_2$/PMMA membrane devices on chip #182.2 with various membrane diameters. Images of 17 different chips are shown in Figure S1.

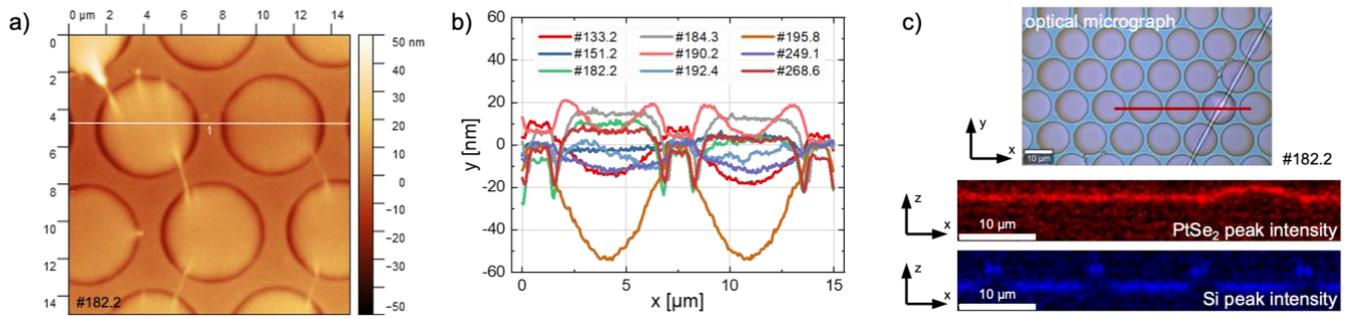

Figure 2. (a) AFM height map of a device with suspended PtSe$_2$/PMMA membranes. AFM scan images of 8 additional chips are shown in Figure S2. (b) Height profiles extracted from the AFM height maps of the 9 chips (Figure S2). (c) Raman depth scan along a line on a device with PtSe$_2$/PMMA membranes. The scan line is shown in red on the microscope image. The intensities of the main PtSe$_2$ peak (E$_g$ peak at ~178 rel. cm$^{-1}$) and the Si peak (at 520 rel. cm$^{-1}$) are shown in the two maps in red and blue.

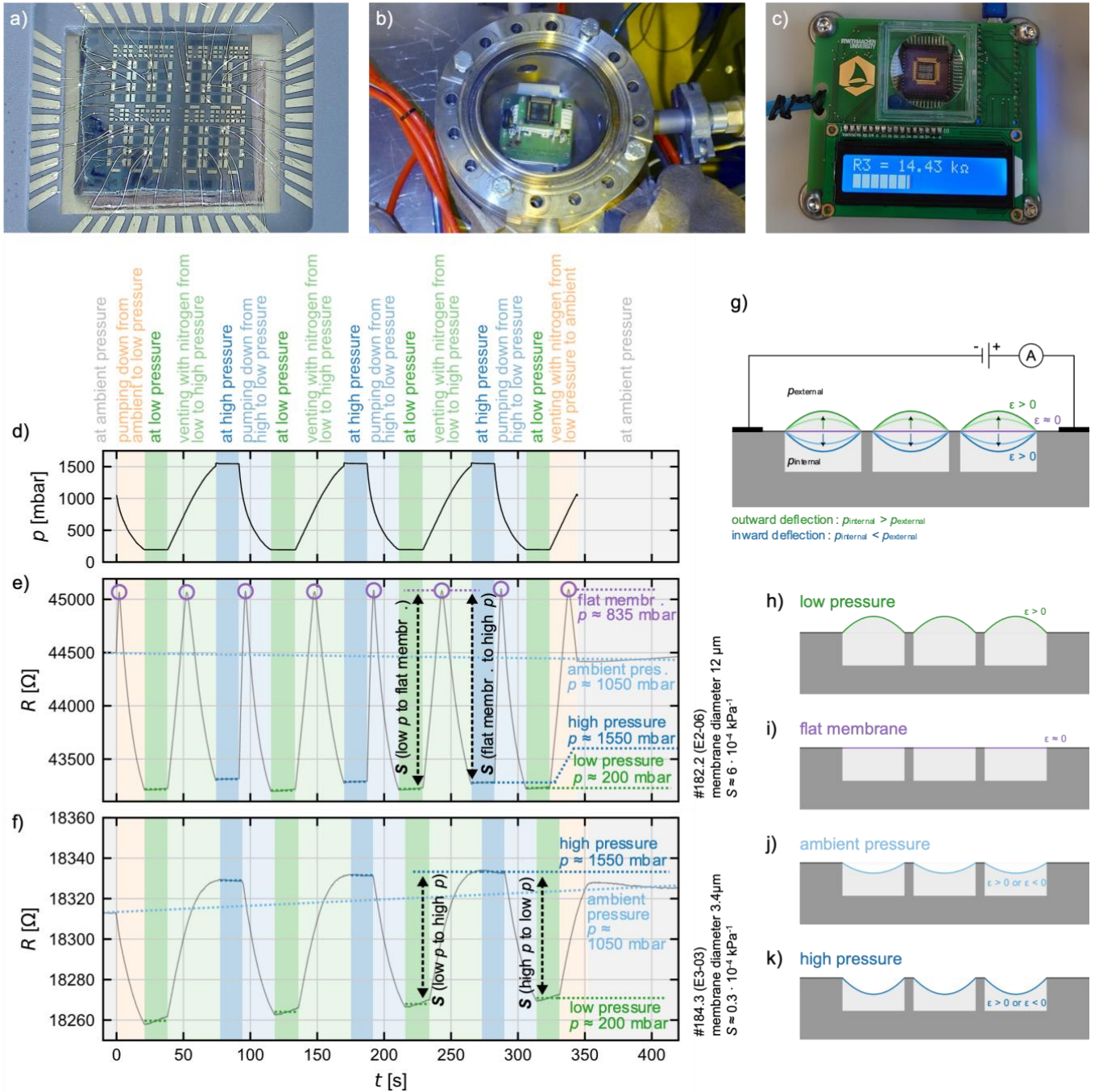

Figure 3. Measurements of the PtSe$_2$/PMMA pressure sensors. (a) A PtSe$_2$ pressure sensor chip wire bonded into a chip carrier. (b-c) Photographs of the pressure chamber for sensor characterization and the pressure sensor demonstrator, respectively. (d) An example measurement curve of the change in pressure over time as recorded by a commercial reference sensor. (e) An example measurement curve of a PtSe$_2$/PMMA pressure sensor with a large membrane diameter (12 µm). The resistance is maximized when the membranes are flat, as indicated by the purple circles. For both the low- and high-pressure cases, tensile strain dominates along the PtSe$_2$, and the resistance decreases. The sensitivity between the resistance maxima of the flat membranes and the levels of low and high pressure was extracted as indicated. The device has a high sensitivity of approx. $6 \cdot 10^{-4}$ kPa$^{-1}$. (f) Exemplary measurement curve of a PtSe$_2$/PMMA pressure sensor with a small membrane diameter

(3.4 µm). Owing to the combination of tensile and compressive strains, high pressure leads to an increase in the measured resistance, whereas low pressure leads to a decrease. The sensitivity is extracted as indicated between the levels of high and low pressure. The device has a low sensitivity of approx. $0.3 \cdot 10^{-4}$ kPa$^{-1}$. (g) Schematic of the working principle of the piezoresistive pressure sensor. (h-k) Sketched membrane deflection for the cases of low pressure, a flat membrane, ambient pressure, and high pressure, as indicated in the measurement curves in (e) and (f).

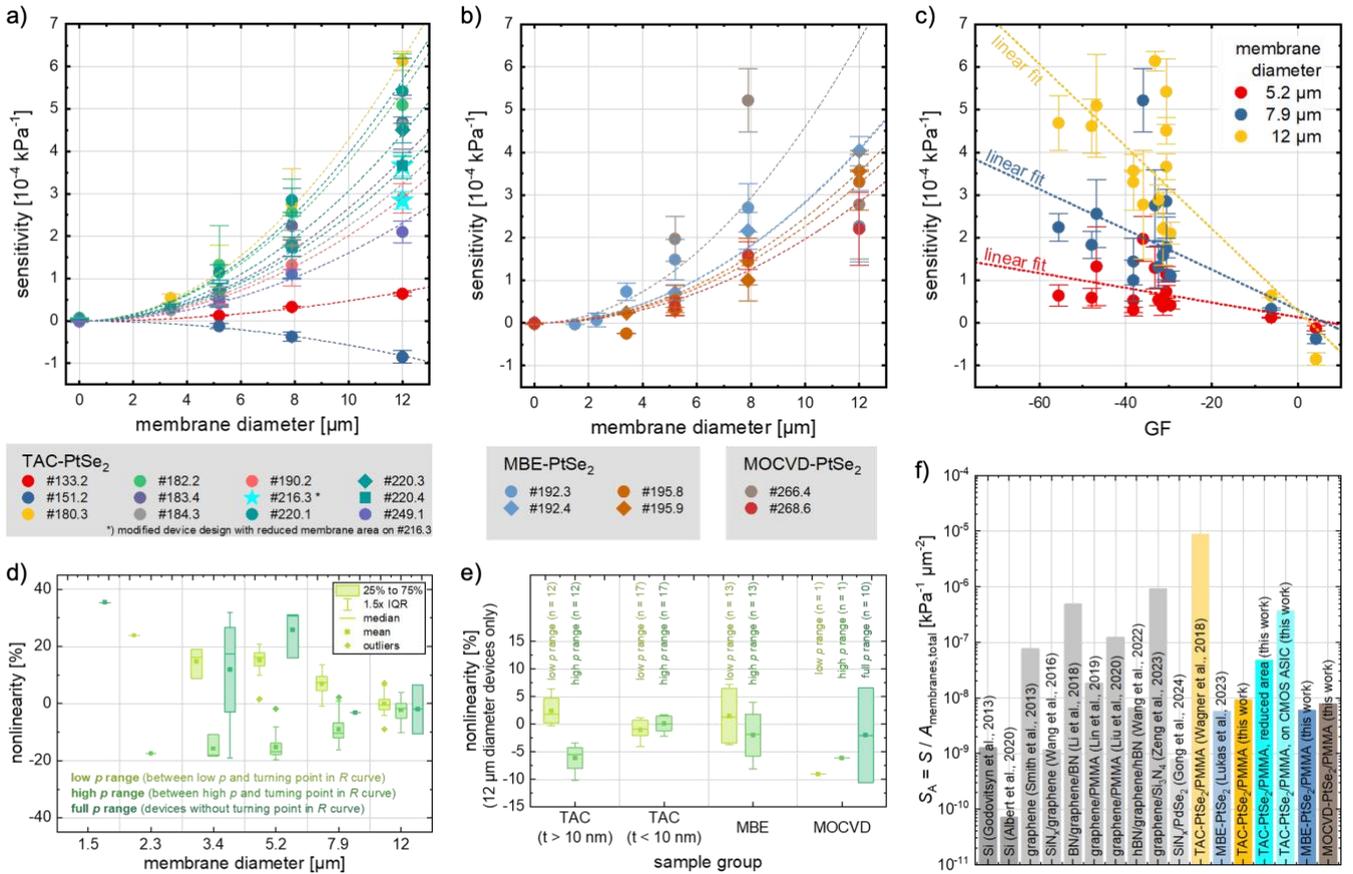

*Figure 4. (a-b) Sensitivity of the PtSe$_2$/PMMA pressure sensors vs. the membrane diameter of the respective device, plotted for 18 different chips. Parabolic fits (without offset and without linear terms) are shown for each chip. (c) Dependence of the sensitivity on the piezoresistive gauge factor (GF) of PtSe$_2$. The graph includes the data points from all chips from graphs (a) and (b), categorized by membrane diameter. Linear fits for the three different membrane diameters are plotted. In (a), (b), and (c), the error bars represent the range of extracted sensitivity values for the individual devices. (d) Nonlinearity (NL) dependence on the membrane diameter. Depending on the sensor characteristics, the NLs extracted from the full pressure range or the two high- and low-pressure ranges are shown. (e) Boxplot of NLs extracted from all devices with the same membrane diameter of d = 12 μm, grouped by material growth and thickness range for TAC-PtSe$_2$. (f) Comparison of the mean sensitivity normalized to the membrane area ($S_A$) for various piezoresistive pressure sensors (see also Table 2).*

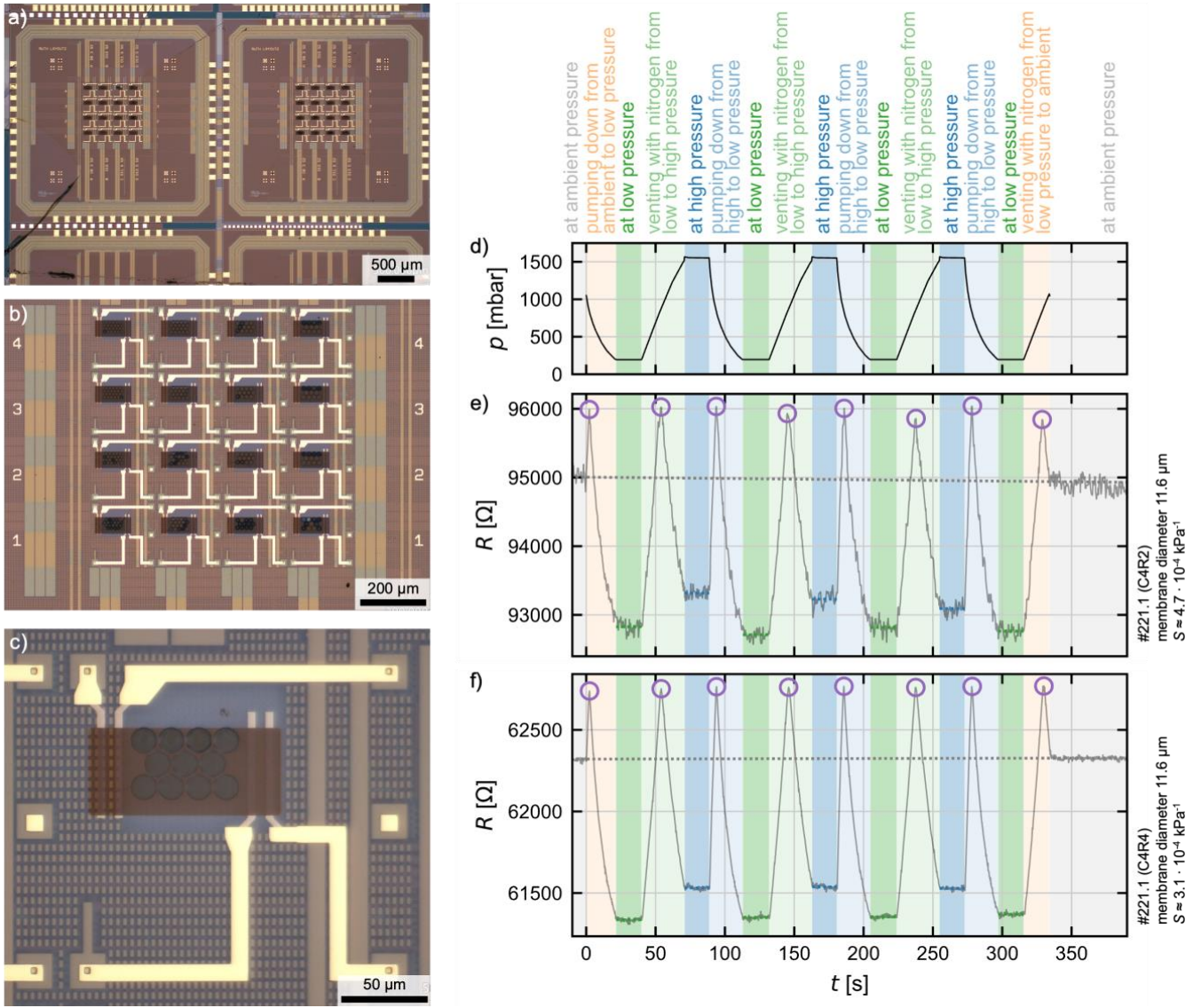

*Figure 5. (a-c) Optical microscope images of PtSe$_2$/PMMA pressure sensors on CMOS ASIC substrates. (d) Reference curve of the chamber pressure during measurement. (e-f) Measurement curves of two PtSe$_2$/PMMA pressure sensors on CMOS ASIC substrates.*

# Supplementary Information for

# Piezoresistive PtSe$_2$ pressure sensors with reliable high sensitivity and their integration into CMOS ASIC substrates

Sebastian Lukas[1], Nico Rademacher[1,2], Sofía Cruces[1], Michael Gross[1], Eva Desgué[3], Stefan Heiserer[4], Nikolas Dominik[4], Maximilian Prechtl[4], Oliver Hartwig[4], Cormac Ó Coileáin[4], Tanja Stimpel-Lindner[4], Pierre Legagneux[3], Arto Rantala[5], Juha-Matti Saari[5], Miika Soikkeli[5], Georg S. Duesberg[4], Max C. Lemme[1,2]

[1]Chair of Electronic Devices, RWTH Aachen University, Otto-Blumenthal-Str. 25, 52074 Aachen, Germany

[2]AMO GmbH, Advanced Microelectronic Center Aachen, Otto-Blumenthal-Str. 25, 52074 Aachen, Germany

[3]THALES R&T, 1 Av. Augustin Fresnel, 91767 Palaiseau, France

[4]Institute of Physics & SENS Research Centre, University of the Bundeswehr Munich, Werner-Heisenberg-Weg 39, 85577 Neubiberg, Germany

[5]VTT Technical Research Centre of Finland Ltd, P.O. Box 1000, FI-02044 VTT, Espoo, Finland

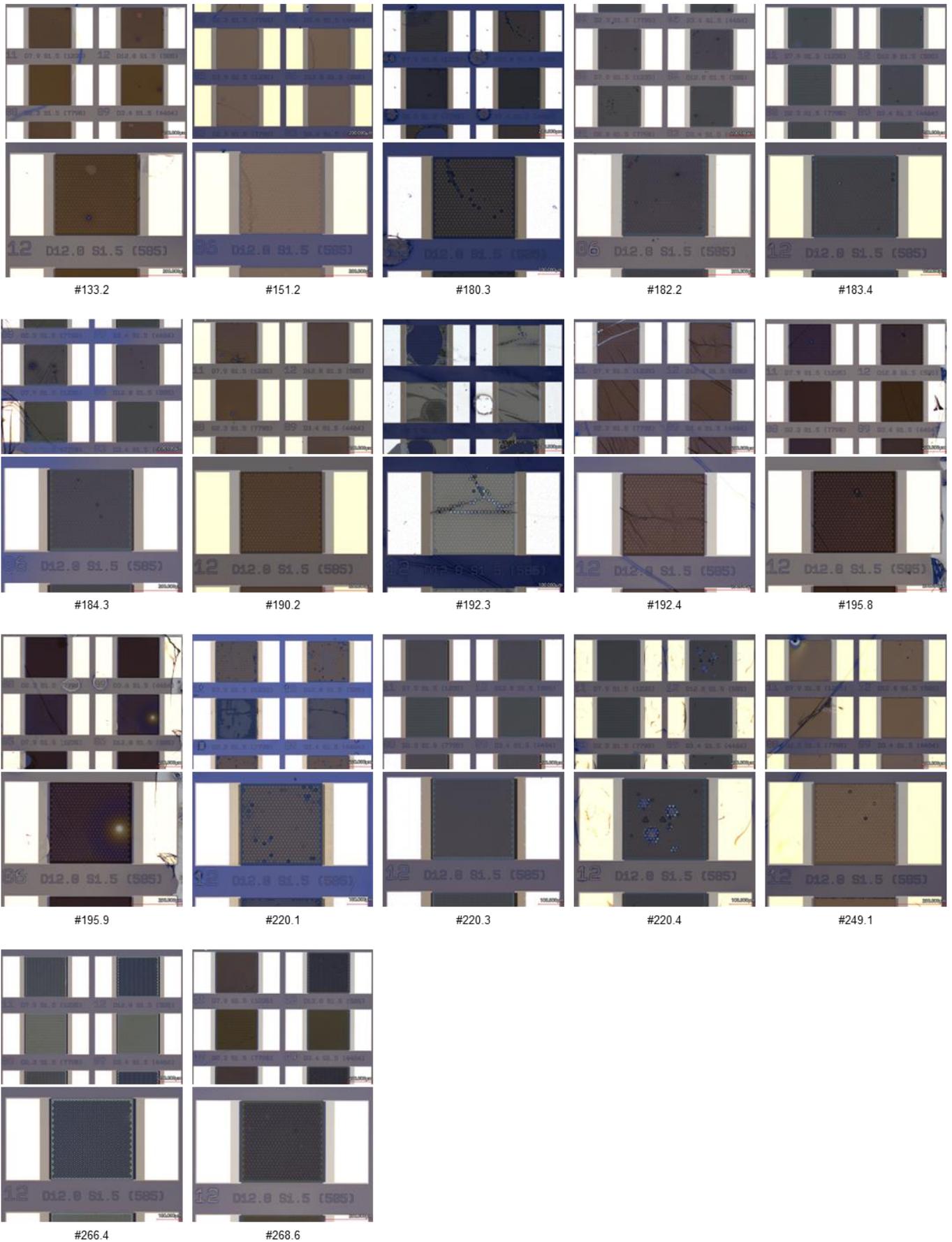

Figure S1. Optical microscopy images of devices on 17 different PtSe$_2$/PMMA pressure sensor chips.

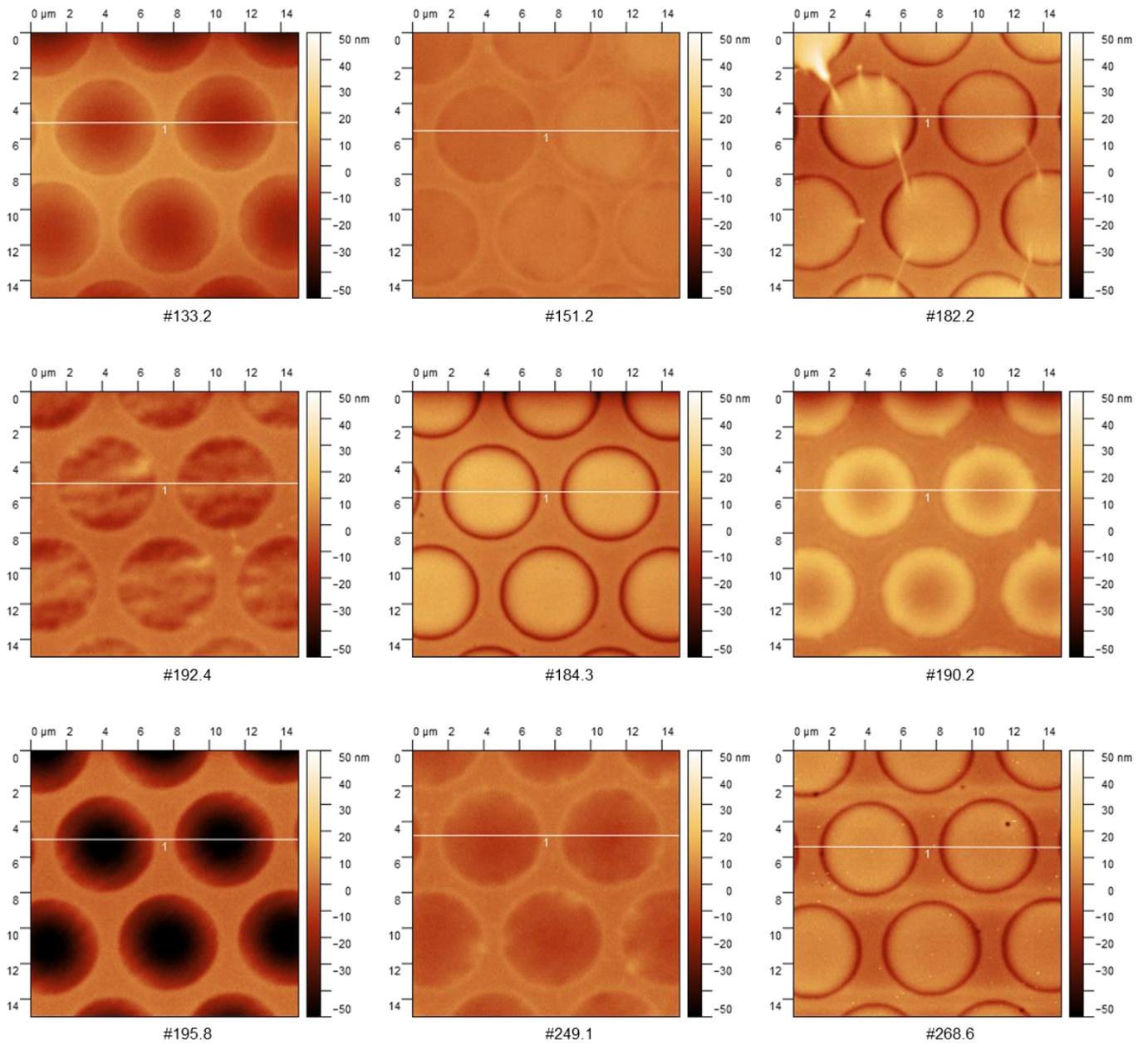

*Figure S2. AFM height maps of 9 different PtSe$_2$/PMMA membrane chips. The profiles extracted along the white horizontal lines are shown in Figure 2b.*

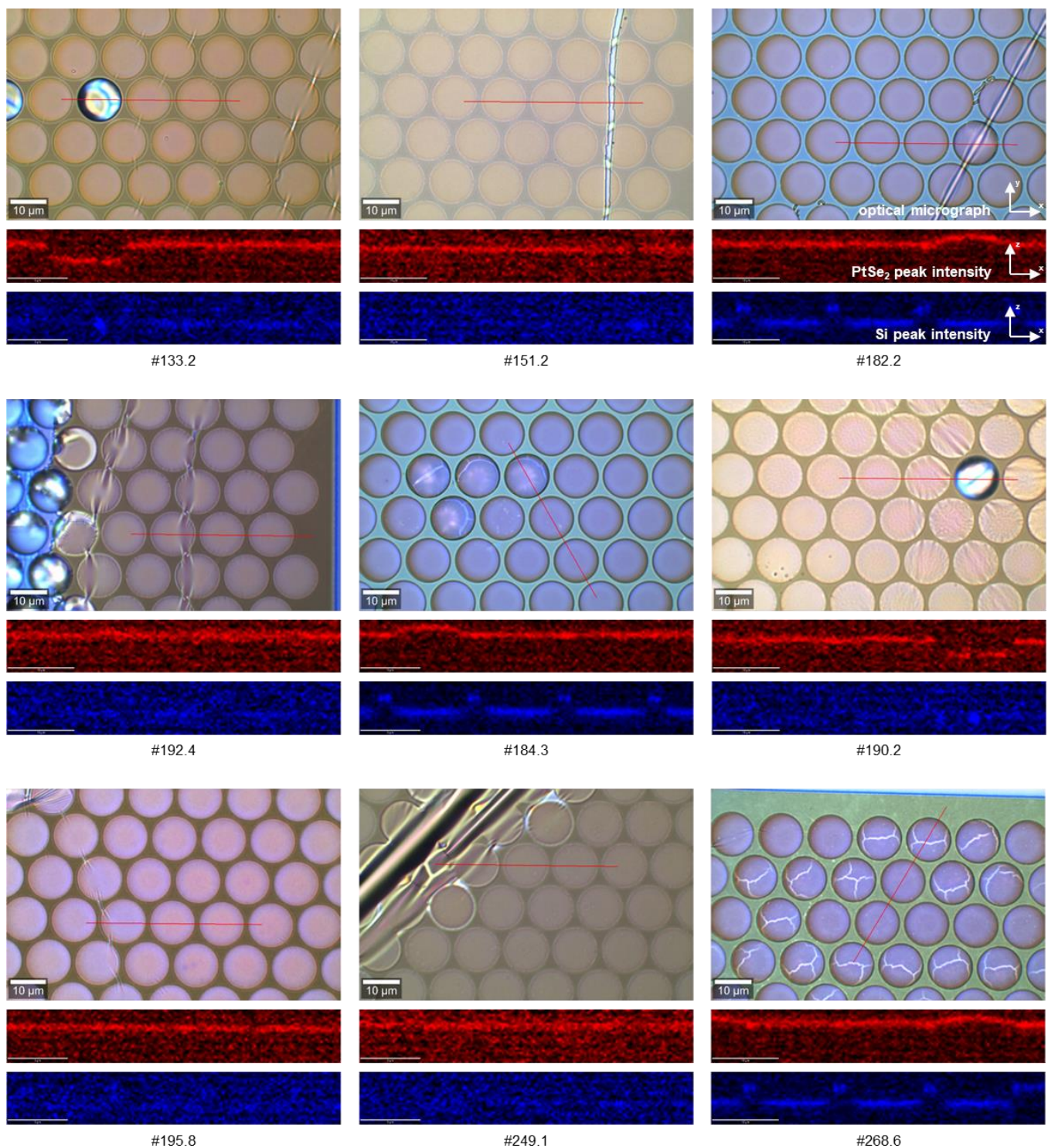

Figure S3. Raman depth scans of 9 different chips. Note that for the Raman scans, an area with a broken membrane (or a membrane with a wrinkle/fold in the PtSe$_2$/PMMA film) was chosen intentionally to demonstrate the capabilities of the method. The broken membranes can be identified in the maps of the PtSe$_2$ peak intensity (E$_g$ peak). In the maps of the Si peak intensity, the profile of the underlying cavities can be seen for those chips where PtSe$_2$ is sufficiently transparent.

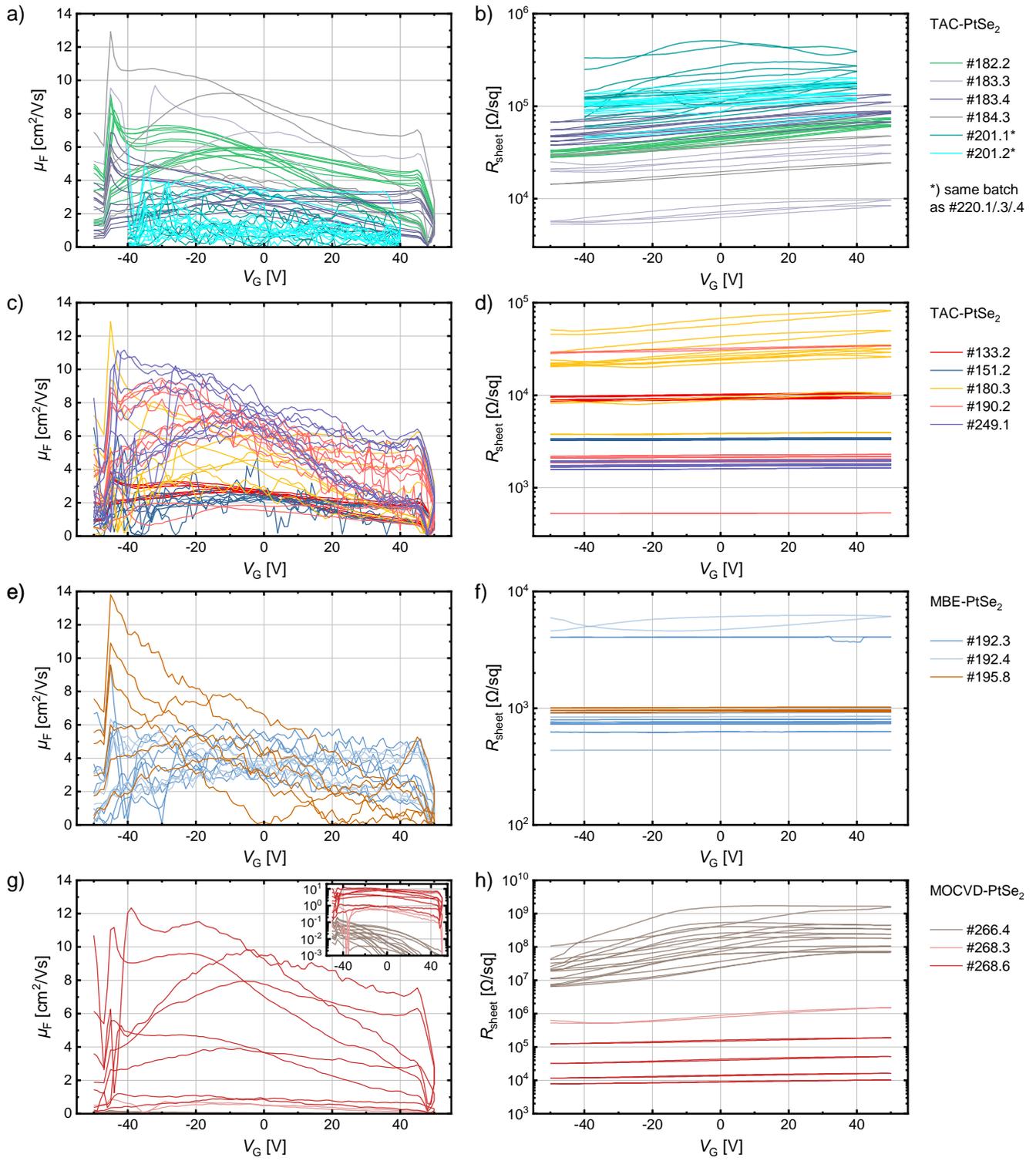

*Figure S4. Four-probe back-gated transfer measurements of PtSe$_2$ reference devices, showing effective charge carrier mobility (left side) and sheet resistance (right side). The chip numbers are indicated to the right.*

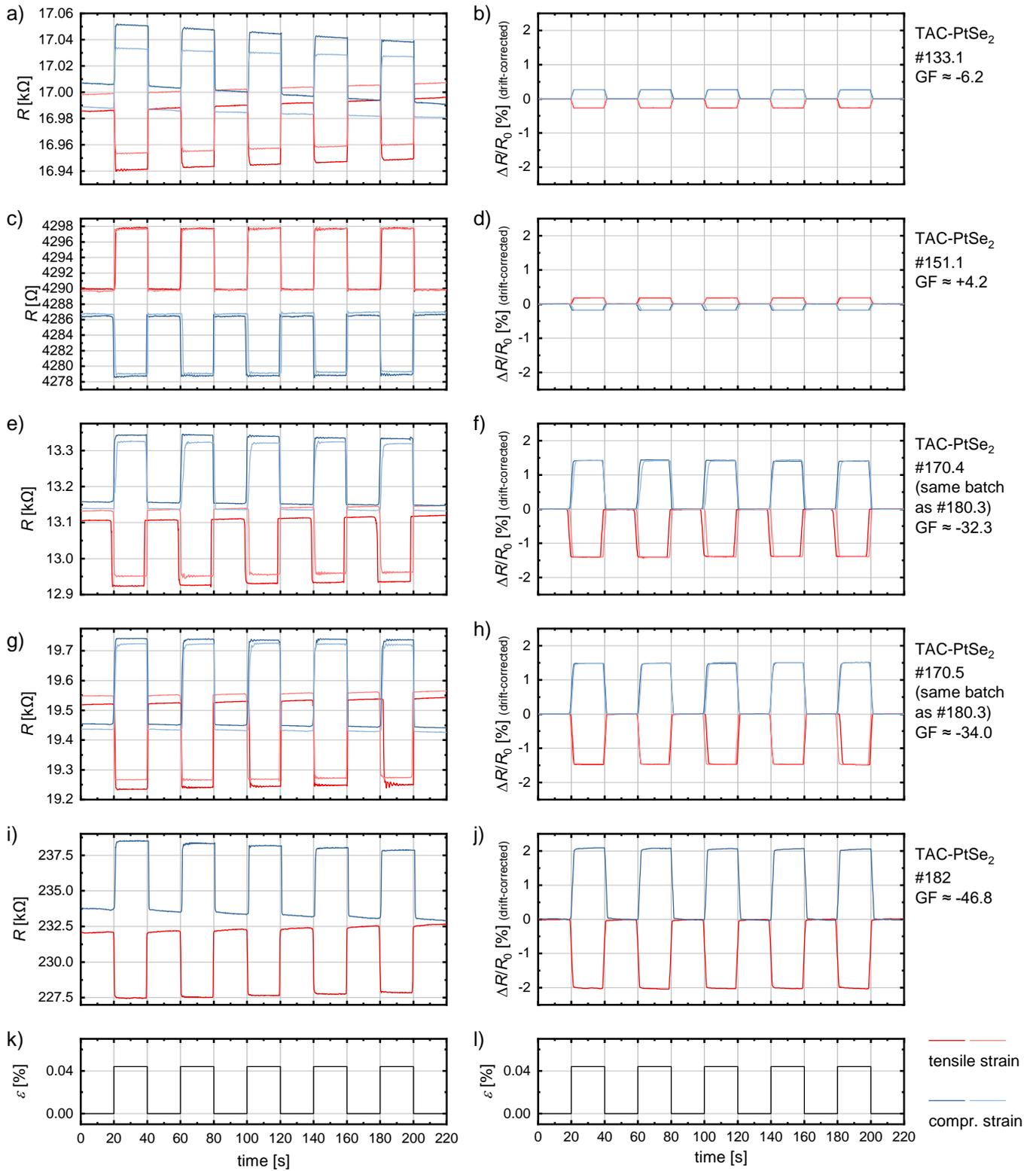

Figure S5. Measurement curves of the PtSe$_2$ resistance R under tensile and compressive strains for extraction of the piezoresistive gauge factor (GF). The absolute and relative changes in R are shown. The sample numbers and extracted GFs are indicated on the right side.

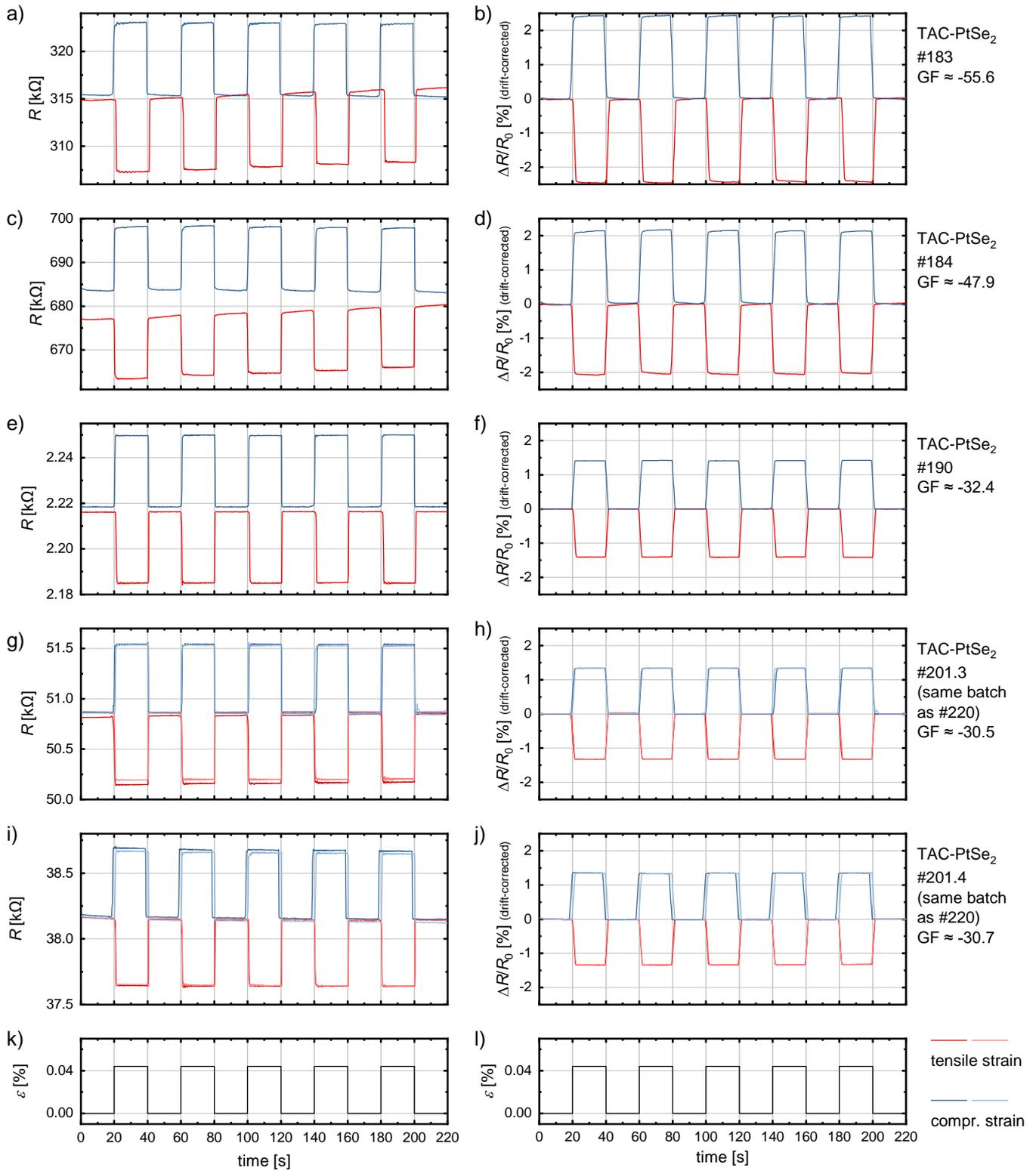

Figure S6. Additional measurement curves of the PtSe$_2$ resistance R under tensile and compressive strains for extraction of the piezoresistive gauge factor (GF). The absolute and relative changes in R are shown. The sample numbers and extracted GFs are indicated on the right side.

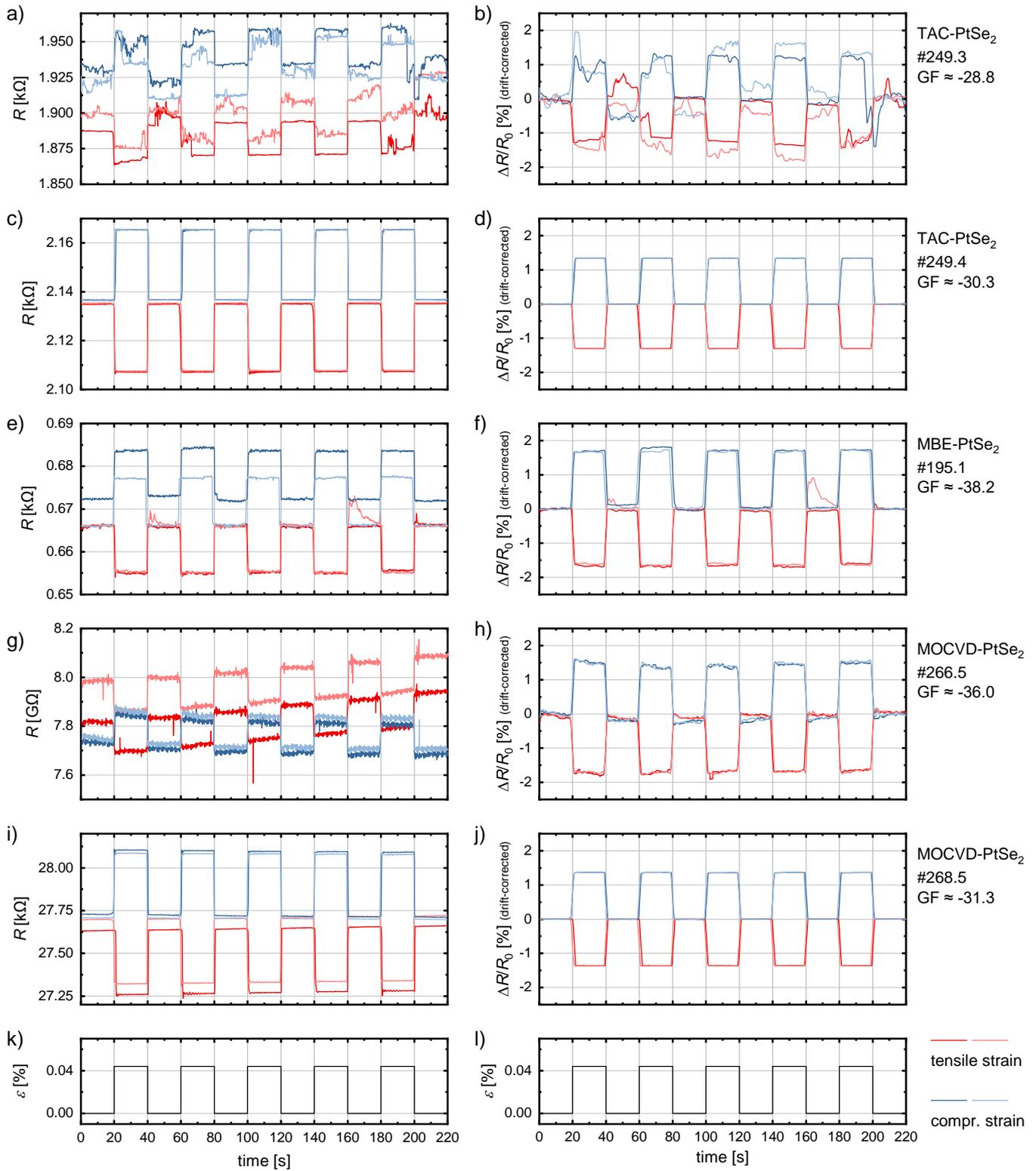

Figure S7. Additional measurement curves of the PtSe$_2$ resistance R under tensile and compressive strains for extraction of the piezoresistive gauge factor (GF). The absolute and relative changes in R are shown. The sample numbers and extracted GFs are indicated on the right side.

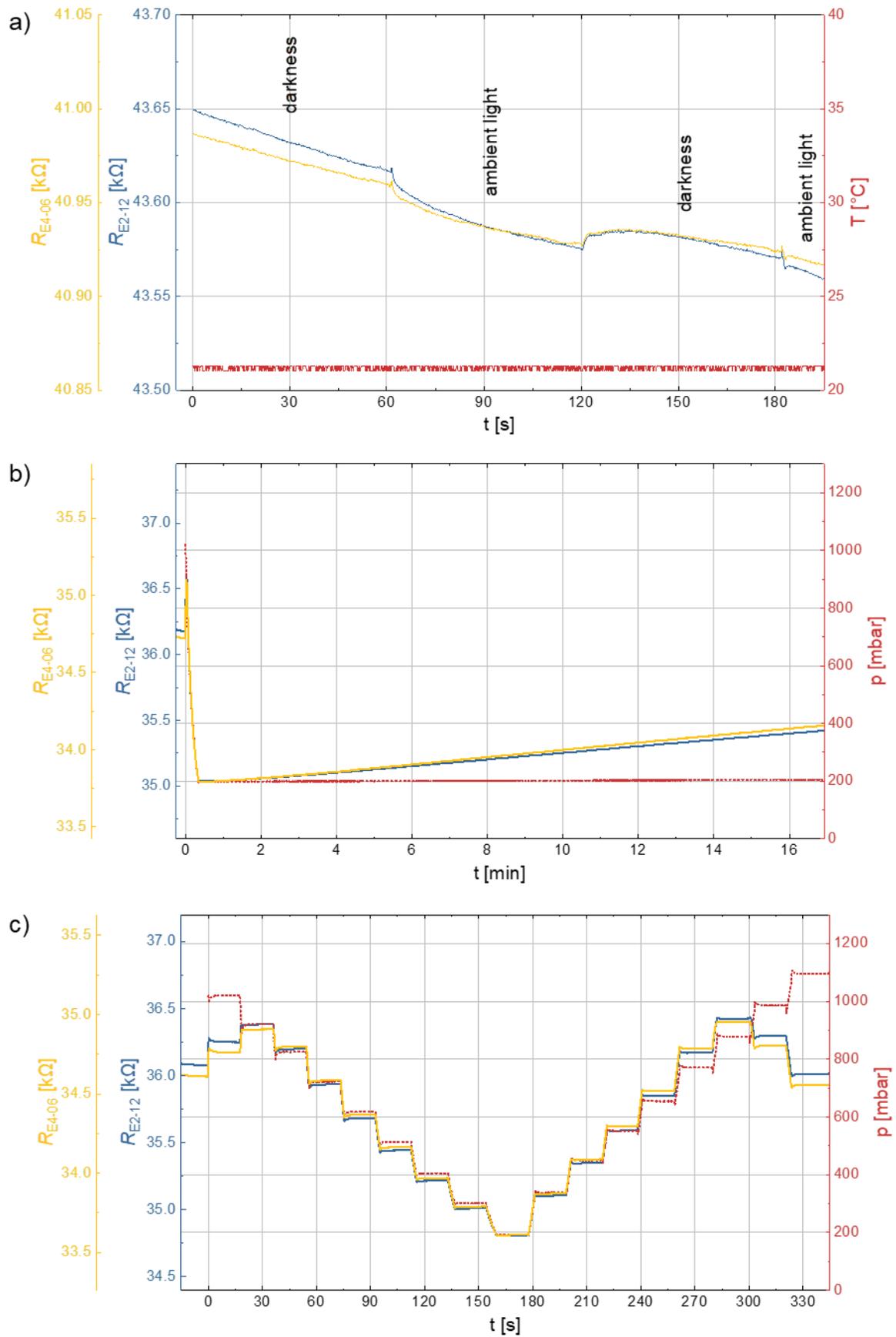

Figure S8. (a) Measurement of two PtSe$_2$/PMMA pressure sensors on chip #220.1 in darkness and under ambient indoor room light conditions. (b) Long-term and (c) pressure-step measurements of two PtSe$_2$/PMMA pressure sensors on chip #220.1.

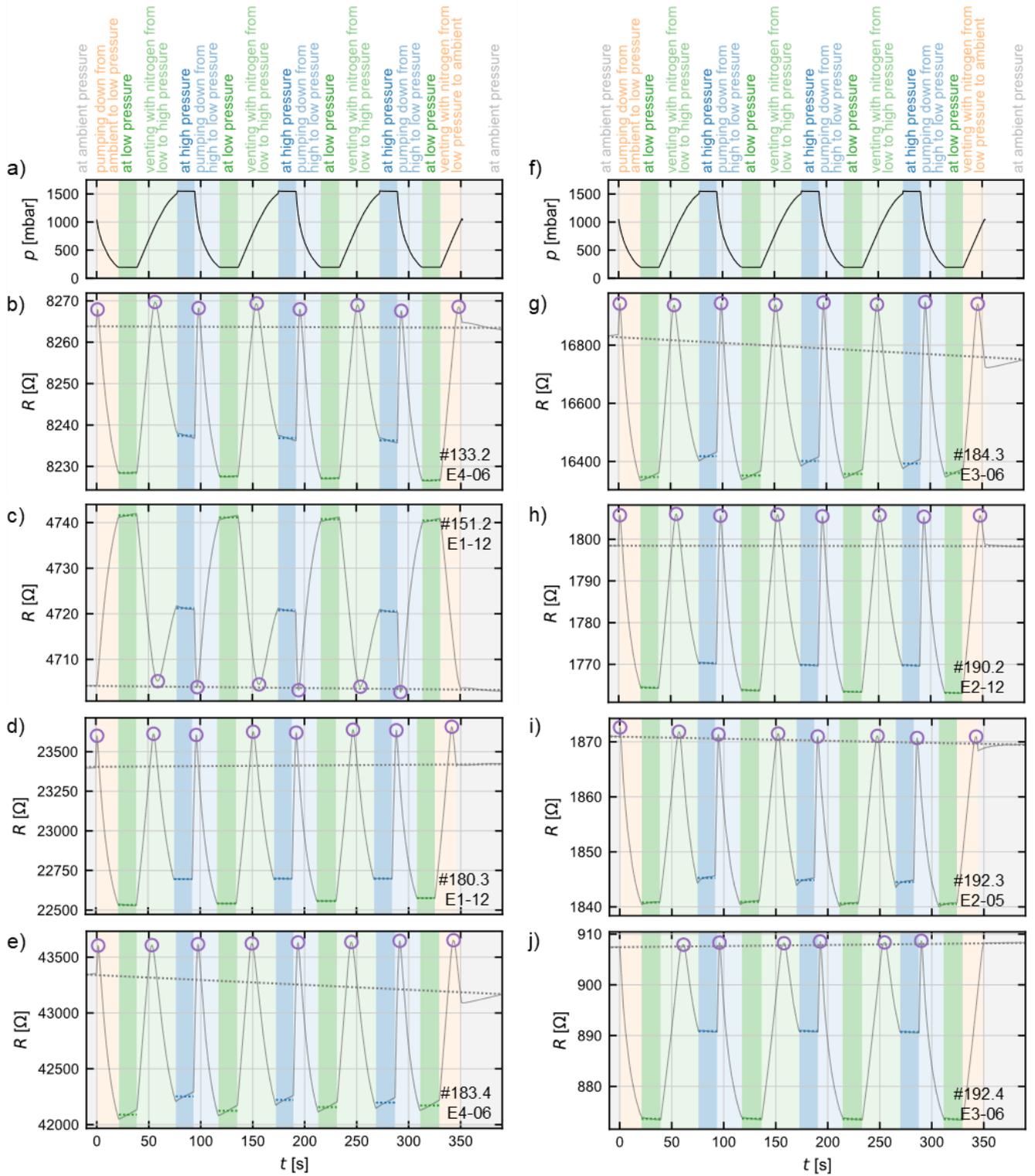

Figure S9. Additional measurement curves of the PtSe$_2$/PMMA pressure sensors. The chip and device numbers are indicated within the panels.

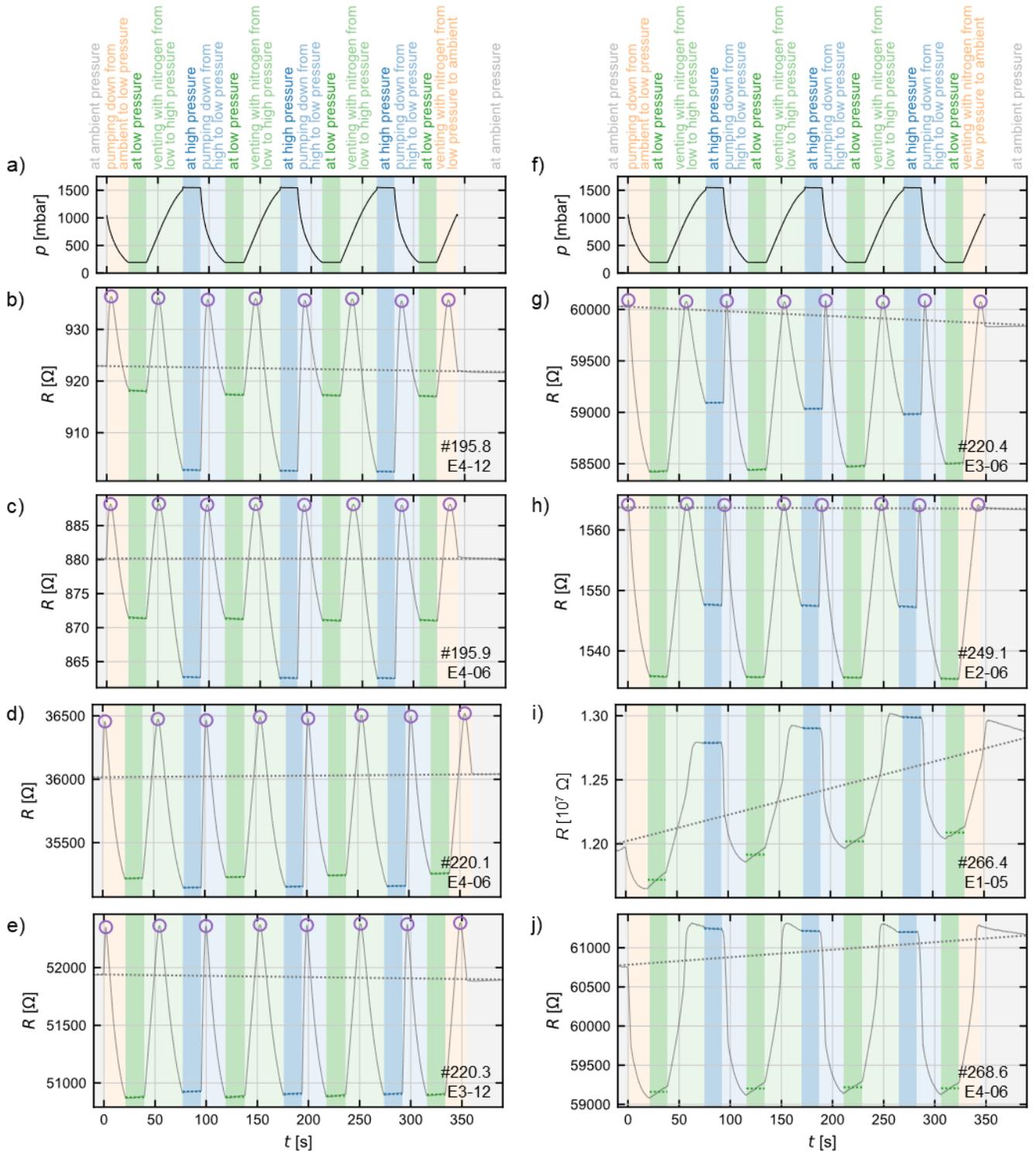

Figure S10. Additional measurement curves of the PtSe$_2$/PMMA pressure sensors. The chip and device numbers are indicated within the panels.

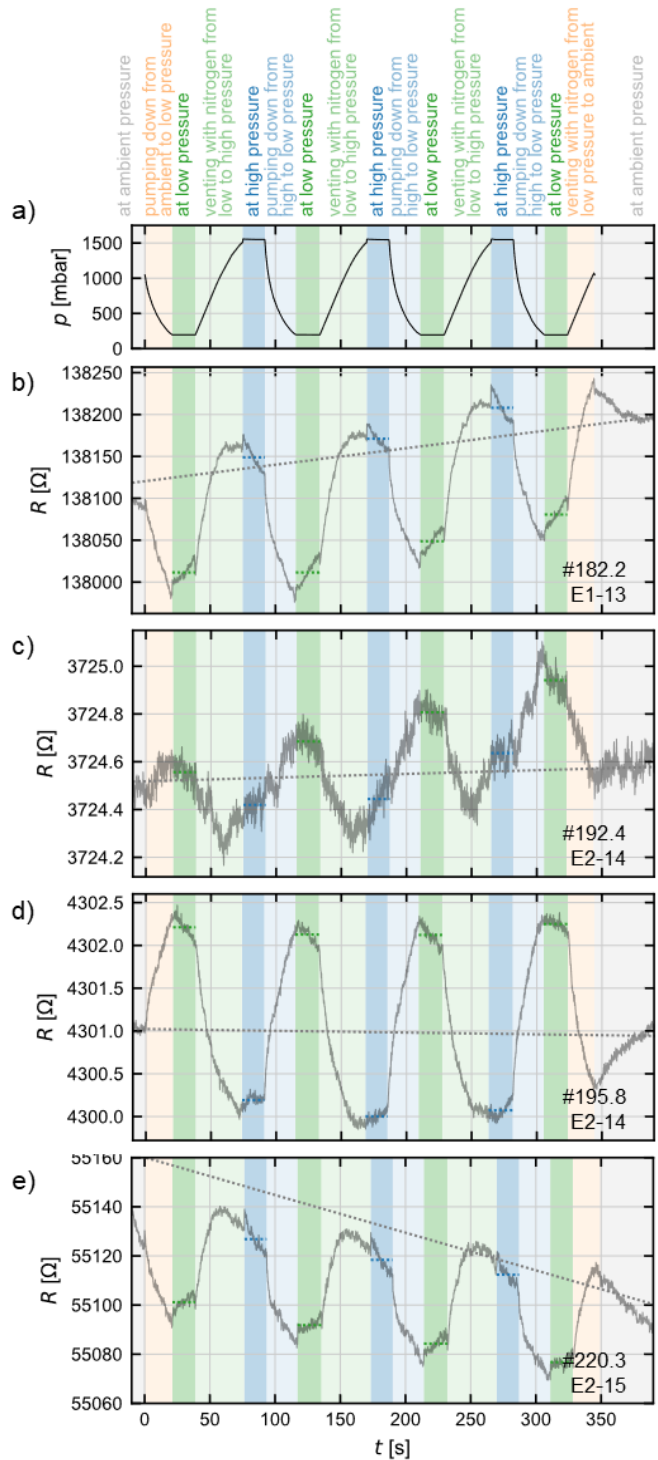

Figure S11. Measurement curves of PtSe$_2$/PMMA reference devices without cavities. Note that the y-axes are significantly enlarged compared with the plots of the respective membrane devices on the same chips (see Figure 3b, Figure S9j, and Figure S10b and e). The sensitivities extracted from the reference devices range from approximately -0.04 to +0.07 · 10$^{-4}$ kPa$^{-1}$, i.e., 2 orders of magnitude lower than the sensitivities extracted from the membrane devices.

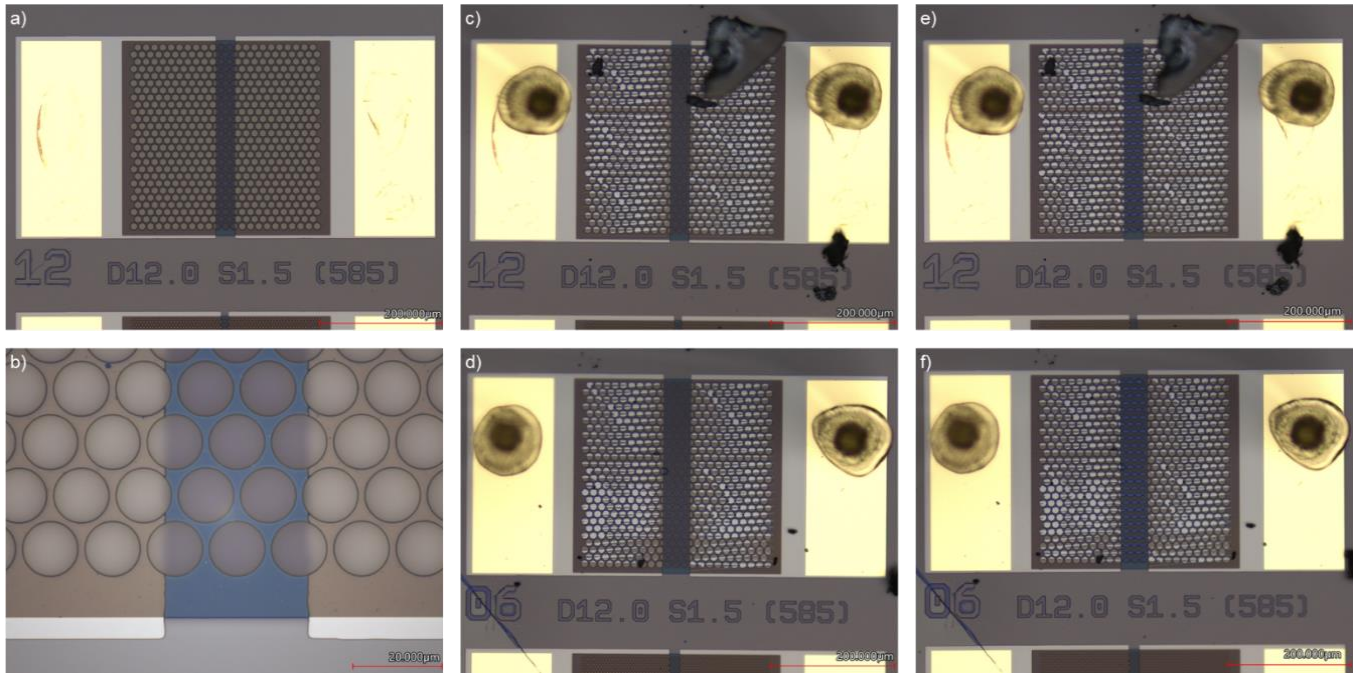

Figure S12. Optical microscope images of devices on chip #216.3 with a reduced active membrane area. (a-b) Device E1-12 after PtSe$_2$/PMMA channel patterning. The Ti/Pd metal contacts cover most of the cavity array except for the narrow gap in the middle. All the PtSe$_2$/PMMA membranes on top of the metal-covered cavities are shorted. (c-d) Devices E1–12 and E1–06 after wire bonding and after the burning of the membranes on top of the metal-covered cavities by a 10 mW laser. The membranes in the narrow active area are still intact. Measurements of the devices in this state were used to evaluate the sensitivity. (e-f) Devices E1–12 and E1–06 after additional burning of the membranes in the active area by a 10 mW laser. Measurements of the devices in this state still revealed device resistances on the same order of magnitude but no response to the change in pressure, confirming that the membranes in the narrow active area generated the sensor signal.

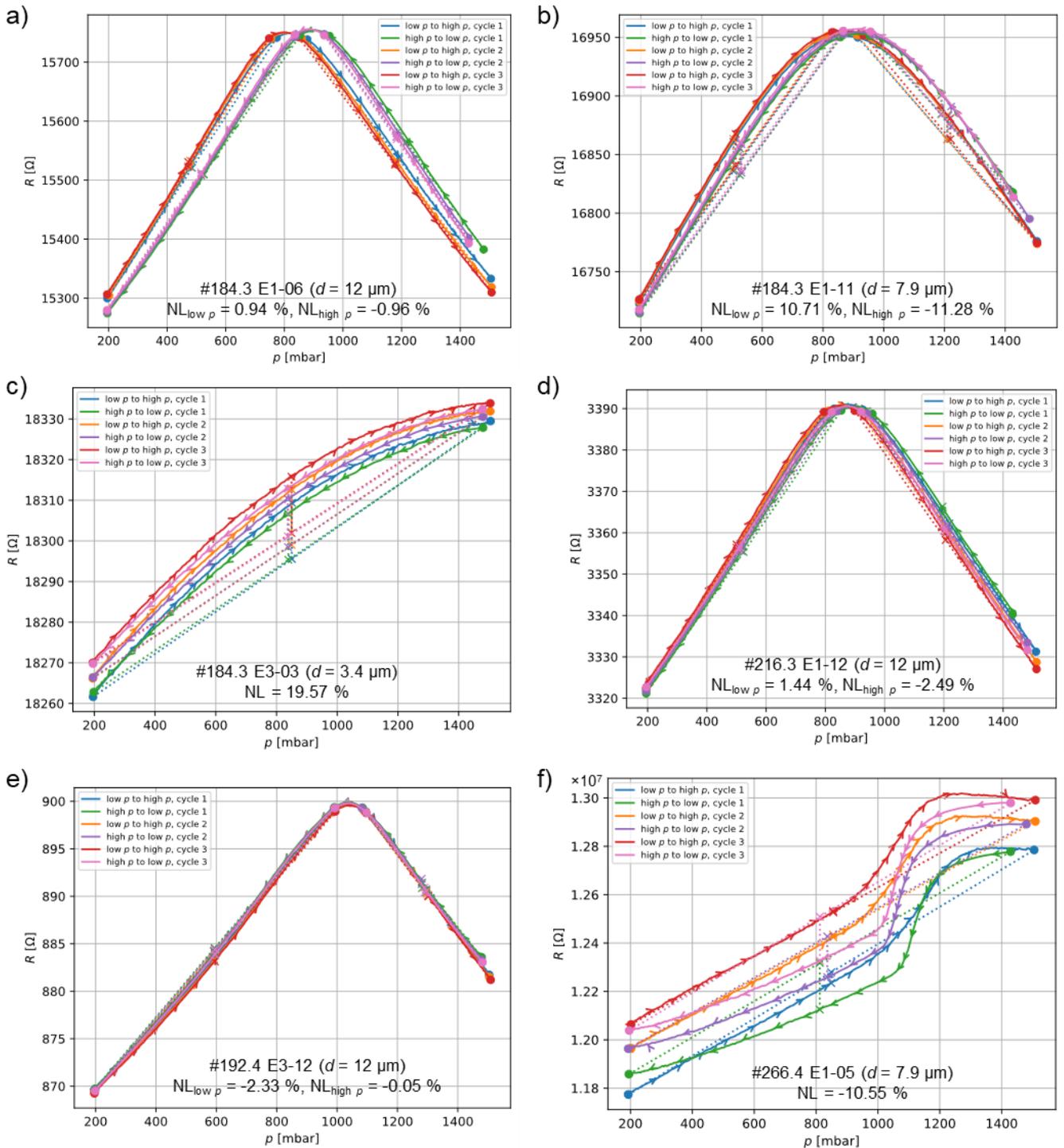

Figure S13. Device resistance versus pressure (R–p) curves of six PtSe$_2$/PMMA pressure sensors from different chips extracted from the time-based measurement curves for the evaluation of nonlinearity (NL). The chip number, membrane diameter, and extracted NL values are noted inside the panels. The NL was extracted from the two ranges to either side of the maximum in the R–p curve (referred to as the low- and high-pressure ranges), with a 50 mbar distance to the maximum. The dotted lines indicate the straight line between $R_{p,min}$ and $R_{p,max}$ (solid circles).

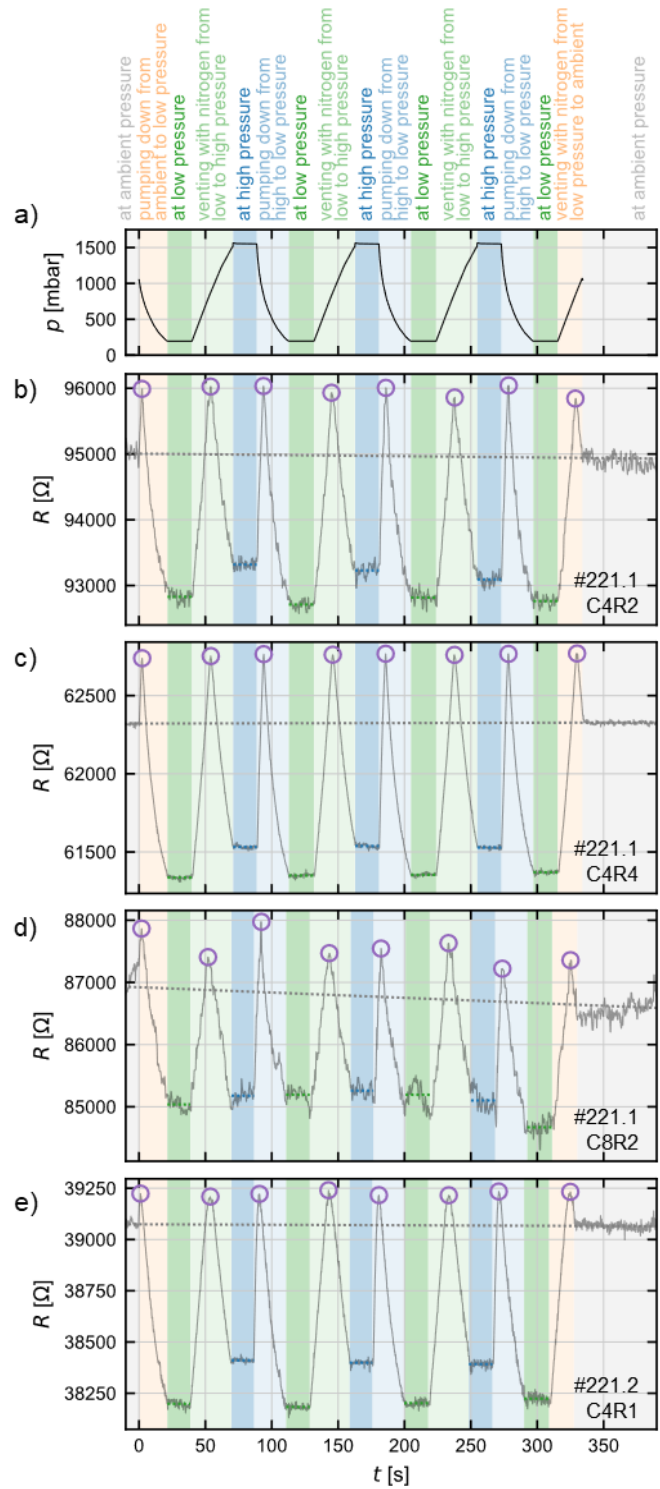

Figure S14. Measurement curves of PtSe$_2$/PMMA pressure sensors on CMOS ASIC substrates. The chip and device numbers are indicated within the panels.

Table S1. Overview of different device dimensions in the main device layout. Each chip (indicated by # throughout the work) featured six different device designs, except for the CMOS ASIC substrates. The device design varies with the membrane diameter. Device labeling was performed according to #[chip number] [die label]-[device number], e.g., #182.2 E2-06, featuring a device "06" with a membrane diameter of 12 µm, according to the table below.

| device numbers | membrane diameter (spacing: 1.5 µm) | membrane count | active area (in case of 100 % intact membrane yield) | |
| --- | --- | --- | --- | --- |
| | | | absolute | relative (total: 320 µm × 320 µm) |
| 01 07 | 1.5 µm | 12,257 | 21,659.9 µm² | 21.2 % |
| 02 08 | 2.3 µm | 7,790 | 32,365.6 µm² | 31.6 % |
| 03 09 | 3.4 µm | 4,464 | 40,529.6 µm² | 39.6 % |
| 04 10 | 5.2 µm | 2,412 | 51,224.0 µm² | 50.0 % |
| 05 11 | 7.9 µm | 1,235 | 60,535.6 µm² | 59.1 % |
| 06 12 | 12 µm | 585 | 66,161.9 µm² | 64.6 % |